\documentclass[11pt]{article}
\usepackage{amssymb}
\usepackage{latexsym,epsfig,amsmath}
\usepackage{epsfig,graphicx, multirow}
\usepackage{amsfonts}
\usepackage{tabularx}
\usepackage{rotating}
\usepackage{harvard}
\usepackage{epstopdf}
\usepackage{booktabs}
\usepackage{rotating}
\usepackage{geometry}
\usepackage{caption}
\usepackage{subcaption}
\usepackage{bbm}
\usepackage{color,soul}
\usepackage{xcolor}

\newcolumntype{z}{>{\centering \arraybackslash}X}
\newtheorem{theorem}{Theorem}

\newtheorem{corollary}{Corollary}

\newtheorem{definition}{Definition}

\newtheorem{assumption}{Assumption}

\begin{document}

\title{Bounds On Treatment Effects On Transitions\thanks{%
We are grateful for helpful suggestions from the editor Han Hong, an associate editor, two anonymous referees,
 John Ham, Per Johansson, Adam Rosen, Michael Svarer, Gerard van den Berg and seminar participants at
IFAU-Uppsala, Uppsala University, University of Aarhus and University of
Mannheim. We acknowledge support from the Jan Wallander and Tom Hedelius
Foundation, NSF grants SES 0819612 and 0819638, from the the ESRC Centre for
Microdata Methods and Practice grant RES-589-28-0001, and from the European
Research Council (ERC) grant ERC-2014-CoG-646917-ROMIA.}}
\author{Johan Vikstr\"om\thanks{%
Corresponding author: IFAU-Uppsala and UCLS at Uppsala University, Box 513,
751 20 Uppsala, Sweden; Email: \texttt{johan.vikstrom@ifau.uu.se}.} \and %
Geert Ridder\thanks{%
Department of Economics, Kaprielian Hall, University of Southern California,
Los Angeles, CA 90089, U.S.A.; Email: \texttt{ridder@usc.edu}.} \and Martin
Weidner\thanks{%
Department of Economics, University College London, Gower Street, London
WC1E 6BT, U.K., and CeMMAP; Email: \texttt{m.weidner@ucl.ac.uk}.}}
\maketitle

\begin{abstract}
\noindent This paper considers the identification of treatment effects on
conditional transition probabilities. We show that even under random
assignment only the instantaneous average treatment effect is point
identified. Since treated and control units drop out at different rates,
randomization only ensures the comparability of treatment and controls at
the time of randomization, so that long-run average treatment effects are
not point identified. Instead we derive informative bounds on these average
treatment effects. Our bounds do not impose (semi)parametric restrictions, for example,
proportional hazards. We also explore various assumptions such as
monotone treatment response, common shocks and positively correlated
outcomes that tighten the bounds. \newline
\end{abstract}

\bigskip

\pagebreak

\section{Introduction}

We consider the effect of an intervention if the outcome is a transition
from an initial to a destination state. The population of interest is a
cohort of units that are in the initial state at the time zero. Treatment is
assigned to a subset of the population either at the time zero or at some
later time. Initially we assume that the treatment assignment is random. One
main point made in this paper is that even if the treatment assignment is random,
only certain average effects of the treatment are point identified. This is
because the random assignment of treatment only ensures comparability of the
treatment and control groups at the time of randomization. At later points
in time treated units with characteristics that interact with the treatment
to increase/decrease the transition probability relative to similar control
units leave the initial state sooner/later than comparable control units, so
that these characteristics are under/over represented among the remaining
treated relative to the remaining controls and this confounds the effect of
the treatment.

The confounding of the treatment effect through selective dropout is usually
referred to as dynamic selection. Existing strategies that deal with dynamic
selection rely heavily on parametric or semi-parametric model restrictions. An example
is the approach of \citeasnoun{Abbring2003} who use the Mixed Proportional
Hazard (MPH) model (their analysis is generalized to a multistate model in
Abbring, 2008). In this model, the instantaneous transition or hazard rate is
written as the product of a time effect, the effect of the intervention and
an unobservable individual effect. As shown by \citeasnoun{389}, the MPH
model is nonparametrically identified, so that if the multiplicative
structure is maintained, identification does not rely on arbitrary
functional form or distributional assumptions beyond the assumed
multiplicative specification. A second example is the approach of %
\citeasnoun{Heckman2007} who start from a threshold crossing model for
transition probabilities. Again they establish semi-parametric
identification, although their model requires the presence of additional
covariates, besides the treatment indicator, that are independent of
unobservable errors and have large support.

In this paper, we ask what can be identified if the identifying assumptions
of the semi-parametric models do not hold. We show that, because of dynamic
selection, we cannot point identify most
average treatment effects of interest even under random assignment. However, we derive bounds on
non-point-identified treatment effects, and show under what conditions they
are informative. Our bounds are general, since beyond random assignment, we
make no assumptions on functional form and additional covariates, and we
allow for arbitrary heterogenous treatment effects as well as arbitrary
unobserved heterogeneity. The bounds can also be applied if the treatment
assignment is unconfounded by creating bounds conditional on the covariates
(or the propensity score) that are averaged over the distribution of these
covariates (or the propensity score).

Besides these general bounds, we derive bounds under additional (weak)
assumptions like monotone treatment response and positively correlated
outcomes. We relate these assumptions to the assumptions made in the MPH
model and to assumptions often made in discrete duration models and
structural models. The additional assumptions often tighten the bounds
considerably. We also discuss how to apply our various identification
results to construct asymptotically valid confidence intervals for the
respective treatment effects.

There are many applications in which we are interested in the effect of an
intervention on transition probabilities/rates. The \citeasnoun{Cox1972}
partial likelihood estimator is routinely used to estimate the effect of an
intervention on the survival rate of subjects. Transition models are used in
several fields. \citeasnoun{VandenBerg2001} surveys the models used and
their applications. These models have also been used to study the effect of
interventions on transitions. Examples are \citeasnoun{Ridder1986}, %
\citeasnoun{Card1988}, Bonnal et al. (2007), \citeasnoun{Gritz1993}, %
\citeasnoun{Ham1996}, \citeasnoun{Abbring2003}, and \citeasnoun{Heckman2007}%
. A survey of models for dynamic treatment effects can be found in %
\citeasnoun{Abbring2007}.

\nocite{Bonnal1997}

An alternative to the effect of a treatment on the transition rate is its
effect on the cdf of the time to transition or its inverse, the quantile
function. This avoids the problem of dynamic selection. From the effect on
the cdf we can recover the effect on the average duration, but we cannot
obtain the effect on the conditional transition probabilities, so that the
effect on the cdf is not informative on the evolution of the treatment
effect over time. This is a limitation since there are good reasons as to why we
should be interested in the effect of an intervention on the conditional
transition probability or the transition/hazard rate. One important reason
is the close link between the hazard rate and economic theory (%
\citeasnoun{VandenBerg2001}). Economic theory often predicts how the hazard
rate changes over time. For example, in the application to a job bonus
experiment considered in this paper, labor supply and search models predict
that being eligible for a bonus if a job is found, increases the hazard rate
from unemployment to employment. According to these models there is a
positive effect only during the eligibility period, and the effect increases
shortly before the end of the eligibility period. The timing of this
increase depends on the arrival rate of job offers and is an indication of
the control that the unemployed has over his/her re-employment time. Any
such control has important policy implications. This can only be analyzed by
considering how the effect on the hazard rate changes over time.

The evolution of the treatment effect over time is of key interest in
different fields. For instance, consider two medical treatments that have
the same effect on the average survival time. However, for one treatment the
effect does not change over time while for the other the survival rate is
initially low, e.g., due to side effects of the treatment, while after that
initial period the survival rate is much higher. As another example,
research on the effects of active labor market policies often documents a
large negative lock-in effect and a later positive effect once the program
has been completed, see e.g. the survey by Kluve et al. (2007).

\nocite{Kluve2007}

We apply our bounds and confidence intervals to data from a job bonus
experiment previously analyzed by Meyer (1996) among others. As discussed
above economic theory has specific predictions for the dynamic effect of a
re-employment bonus with a finite eligibility period. Meyer (1996) estimates
these dynamic effects using an MPH model. We study what can be identified if
we rely solely on random assignment and some additional (weak) assumptions.

In section \ref{s:param} we define the treatment effects that are relevant
if the outcome is a transition. Section \ref{s:bounds} discusses their point
or set identification in the case that the treatment is randomly assigned.
This requires us to be precise on what we mean by random assignment in this
setting. In section \ref{s:additional} we explore additional assumptions
that tighten the bounds. In section \ref{s:inference} we derive the
confidence intervals. Section \ref{s:application} illustrates the bounds for
the job bonus experiment. Section \ref{s:conclusions} concludes.


\section{Setup}

\label{s:param}

\subsection{Motivating example}

In this paper we consider identification of the effect of a treatment on the
conditional transition probability, usually referred to as the transition
rate or the hazard rate. Effects on transition rates are important in many
applications. The Illinois job-bonus experiment that we re-consider in the
application in this paper is one example. The experiment that was conducted
between mid-1984 and mid-1985 paid re-employment bonuses to unemployed
individuals in the randomized treatment group who found employment within
the first 11 weeks of unemployment. The fact that the bonus is only paid
during the first 11 weeks has several interesting implications. Standard
labor supply and search models predict that being eligible for the bonus
should increase the transition rate from unemployment to employment during
the 11 week eligibility period, but should have no effect after the end of
the eligibility period. Another prediction is that the transition rate
should increase shortly before the end of the eligibility period, as the
unemployed run out of time to collect the bonus. These theoretical
predictions can only be studied by examining how the effect of the job-bonus
varies with time in unemployment, that is by studying the effect on the
transition rate during the eligibility period, shortly before the end of the
eligibility period and after the end of the eligibility period. Effects on
the transition rate are also relevant in many other applications, including
evaluations of medical treatments and active labor market policies.

The job-bonus experiment includes random treatment assignment, which ensures
comparability of the treatment group and the control group at the time of
randomization. At later time points some unemployed individuals have found a
job, and this creates dynamic selection, that even under the initial random
assignment might confound the comparability of the treatment and control
groups. This is most easily seen if the fraction that has found a job
differs between the two groups, and if those who have found a job have more
favourable characteristics than those who remain unemployed. Under these
conditions the remaining individuals in the treatment group will be
negatively (positively) selected if the fraction remaining in unemployment
is lower (higher) in the treatment group than in the control group.
Moreover, even if the fraction still unemployed is the same in the treatment
group and the control group we might still face a selection problem. In the
job-bonus experiment, it could, for instance, be the case that individuals
that respond to the bonus come from different parts of the ability
distribution compared to those who find a job without the bonus. The
implication of this is that the ability distribution differs between the
treatment and the control groups, even if the fraction that has found a job
is the same in the two groups. All this constitutes the dynamic selection
problem that is addressed in this paper.

Previous studies that deal with the dynamic selection problem have mostly
used parametric and semi-parametric models. For instance, %
\citeasnoun{Meyer1996} uses a proportional hazard (PH) model to study how
the effect of the job-bonus experiment considered in this paper varies
before and after the 11 week eligibility period. A more general alternative
to the PH model is to use a Mixed Proportional Hazard (MPH) model. In this
model the instantaneous transition or hazard rate is written as the product
of a time effect, the effect of the intervention and an unobservable
individual effect. This model, however, imposes a multiplicative structure,
a homogeneous treatment effect as well as other restrictions. In this paper
we instead consider what can be identified if we rely solely on random
assignment and do not impose the parametric restrictions that are implicit
in the MPH model and other parametric and semi-parametric models.


\subsection{Average treatment effect on transitions}

We discuss the definition and identification of treatment effects on
transition rates in discrete time with transitions occurring at times $%
t=1,2,\ldots $.\footnote{%
The definition of causal effects in continuous time adds technical problems
(see e.g. \citeasnoun{Gill2001}) that would distract from the conceptual
issues.} We assume that treatment is assigned at the beginning of the first
period and that each unit is either always treated or always non-treated. In
section \ref{s:arbitrary} we generalize these results by allowing the
treatment to start in any time period. Let the potential outcome $Y_{t}^{1}$
be the indicator of a transition in period $t$ if treated and similarly $%
Y_{t}^{0}$ be the potential outcome if non-treated.

In any definition of the causal effect of a treatment on the transition rate
we must account for the dynamic selection that was discussed in the previous
subsection. If we do not specify a model for the transition rate we need to
find another way to maintain the comparability of the treatment and control
groups over time. The approach that we take in this paper is to consider
average transition rates where the average is taken over the same population
for both treated and controls (or in general for different treatment arms).
We initially propose to average over the subpopulation of individuals who
would have survived until time $t$ if treated. This is the analogue of the
average effect on the treated considered in the static treatment effect
literature. This leads to the following definition

\begin{definition}
The causal effect on the transition probability of the treated survivors in $%
t$ is the Average Treatment Effect on Treated Survivors (ATETS) defined by
\begin{equation*}
\mathrm{ATETS}_{t}=\mathbb{E}\left( Y_{t}^{1}|Y_{t-1}^{1}=0,\ldots
,Y_{1}^{1}=0\right) -\mathbb{E}\left( Y_{t}^{0}|Y_{t-1}^{1}=0,\ldots
,Y_{1}^{1}=0\right).
\end{equation*}
\end{definition}

The differential selection only starts after the first period and the $%
\mathrm{ATETS}_{t}$ controls for that by comparing the transition rates for
individuals with a common survival experience.\footnote{%
In Appendix C we also consider the average effect for the subpopulation of
individuals who would have survived until $t$ under both treatment and no
treatment$.$}

Note that we are only concerned with the comparability of the treatment and
control groups over the spell, i.e. with the different levels of dynamic
selection in the two groups. If we keep the treatment and control groups
comparable over time, there is still the question of how to interpret the
time path of the average treatment effect over the spell. In this paper we
do not try to decompose this path into the average treatment effect for a
population of unchanging composition and a selection effect relative to this
population. We do not define the treatment effect for this population of
unchanging composition, but rather for a population with a composition that
changes over time due to dynamic selection. The dynamic selection is made
equal in the treatment and control groups, so that the treatment effect is
not confounded by dynamic selection. Again this is analogous to the
difference between the Average Treatment Effect and the Average Treatment
Effect on the Treated in the case of a static treatment effect where the
latter is defined for the population selected for treatment and the
treatment effect is for this selective population.


\section{Bounds on average treatment effects on transitions}

\label{s:bounds}

We now consider identification of the $\mathrm{ATETS}_{t}$ under random
treatment assignment. Let $D$ be the indicator of treatment status, and $%
Y_{t}$ be the observed indicator of a transition in period $t$. The observed
outcomes are related to the potential outcomes by the observation rule\footnote{%
In applications with at most a single transition per individual (where the destination state is absorbing), as in the job-bonus experiment,
we have $\sum_t Y_t \leq 1$, but we still consider future $Y_t$ to be observed, even after the transition occurred.
}
\begin{equation}
Y_{t}=DY_{t}^{1}+(1-D)Y_{t}^{0}.  \label{eqobs}
\end{equation}%
We make the following random assignment assumption

\begin{assumption}[Random assignment of treatment]
~ \label{AssRand3}\
\begin{equation*}
D\, \bot \, \left \{ Y_{t}^{1},\,Y_{t}^{0} \, : \,t=1,2,\ldots \right \} .
\end{equation*}
\end{assumption}

In the first period $t=1$ no dynamic selection has taken place, yet, so
subjects are fully randomized. Under Assumption~\ref{AssRand3} we therefore
have the usual result, for $d \in \{0,1\}$,
\begin{equation}
\mathbb{E}(Y_{1}^{d}) = \mathbb{E}(Y_{1}|D=d) ,  \label{eqY_t1}
\end{equation}
implying that we can point identify the instantaneous treatment effect:
\begin{equation*}
\mathrm{ATETS}_{1}=\mathbb{E}(Y_{1}^{1})-\mathbb{E}(Y_{1}^{0})=\mathbb{E}%
(Y_{1}|D=1)-\mathbb{E}(Y_{1}|D=0)\text{.}
\end{equation*}

Next, we consider the identification of $\mathrm{ATETS}_t$ for $t=2$. We
discuss this two period case in detail, because the main results of this
paper can be understood in this two period setting, where the transition
occurs in period 1, period 2 or after period 2. The two period dynamic
treatment effect is defined by
\begin{equation}
\mathrm{ATETS}_{2} = \mathbb{E}(Y_{2}^{1}|Y_{1}^{1}=0) -\mathbb{E}%
(Y_{2}^{0}|Y_{1}^{1}=0).  \label{DefATETSforT2}
\end{equation}%
Under Assumption~\ref{AssRand3} we again have, for $d \in \{0,1\}$,
\begin{equation*}
\mathbb{E}(Y_{2}^{d}|Y_{1}^{d}=0) = \mathbb{E}(Y_{2}|Y_{1}=0,D=d).
\end{equation*}
Thus, the first term in $\mathrm{ATETS}_{2}$ is point identified from the
data, and we can also point identify $\mathbb{E}(Y_{2}^{0}|Y_{1}^{0}=0)$.
However, in this last expression the conditioning is on the survivors under
non-treatment instead of under treatment, so this is not the second term in $%
\mathrm{ATETS}_{2}$. It turns out that $\mathbb{E}(Y_{2}^{0}|Y_{1}^{1}=0)$
is only partially identified from the data, and the goal in the following is
therefore to derive bounds on this conditional expectation.

For every member of the population we have a vector of four binary potential
outcomes $Y_{1}^{1}$, $Y_{2}^{1}$, $Y_{1}^{0}$, $Y_{2}^{0}$, for which there
are $2^4 = 16$ possible realizations. We denote the probability of $%
(Y_{1}^{1}, Y_{2}^{1}, Y_{1}^{0}, Y_{2}^{0}) = (d_1,d_2,d_3,d_4)$ by $p_{d_1
d_2, d_3 d_4}$. Table~\ref{tab:ProbT2} shows those sixteen population
probabilities, using the two-vector notation $Y=(Y_1,Y_2)$ and $%
Y^d=(Y^d_1,Y^d_2)$. From the data we can identify the transition
probabilities $\Pr \left( Y_1 = 1 \big| D=d\right)$ (transition in $t=1$
under treatment $d$), and $\Pr \left( Y=(0,1) \big| D=d\right)$ (transition
in $t=2$ under treatment $d$), and $\Pr \left( Y=(0,0) \big| D=d\right)$
(transition after $t=2$ under treatment $d$). Those ``observable''
transition probabilities are obtained in Table~\ref{tab:ProbT2} as row- and
column-sums, for example we have $\Pr \left( Y=(0,0) \big| D=1\right) =
p_{00,00} + p_{00,01} + p_{00,10} + p_{00,11}$.

Notice that $Y^d=(1,1)$ is included as a potential outcome here, that is, we
allow for multiple transitions. Multiple transitions cannot occur if the
destination state is absorbing, as in the job bonus experiment.
In that case we know that the probabilities in the last row and column of Table~\ref{tab:ProbT2} are zero, that is,
\begin{equation*}
p_{11,00}=p_{11,01}=p_{11,10}=p_{11,10}=p_{11,11}=p_{00,11}=p_{01,11}=p_{10,11}=0 .
\end{equation*}
This information could sharpen the lower bound on the treatment effect, but
we
will not derive separate bounds for the case of an absorbing destination
state. The bounds for the non-absorbing destination state are conservative
if the destination is indeed absorbing.

With those definitions we obtain\footnote{%
We have
\begin{equation*}
\Pr \left(Y_{2}^0=1 \big| Y_{1}^{1}=0 \right) = \frac{ \Pr \left(Y_{2}^0=1
\, \& \, Y_{1}^{1}=0 \right) } { \Pr \left( Y_{1}^{1}=0 \right) } = \frac{
\Pr \left \{ \left[ Y^0 = (0,1) \; \text{or} \; (1,1) \right] \, \& \, \left[
Y^1 = (0,0) \; \text{or} \; (0,1) \right] \right \} } { \Pr \left( Y_1 =0 %
\big| D=1 \right) } ,
\end{equation*}
where we also used the random assignment assumption.}
\begin{align}
\mathbb{E} \left(Y_{2}^0 \big| Y_{1}^{1}=0 \right) &= \Pr \left(Y_{2}^0=1 %
\big| Y_{1}^{1}=0 \right) = \frac{p_{00,01} + p_{00,11} + p_{01,01} +
p_{01,11} } { \Pr \left( Y_1 =0 \big| D=1 \right) } .  \label{InequT2e1}
\end{align}
The denominator of the last expression is identified from the data. What is
left to do is to provide bounds on the numerator in terms of the six
observable transition probabilities $\Pr \left( Y_1 = 1 \big| D=d\right)$, $%
\Pr \left( Y=(0,1) \big| D=d\right)$, and $\Pr \left( Y=(0,0) \big| %
D=d\right) $, $d \in \{0,1\}$. The four probabilities that enter into this
numerator are underlined in Table~\ref{tab:ProbT2}.

Thus, the question is what values for $p_{00,01} + p_{00,11} + p_{01,01} +
p_{01,11}$ are feasible, subject to the positivity condition $p_{d_1 d_2,
d_3 d_4} \geq 0$ for all $(d_1,d_2,d_3,d_3) \in \{0,1\}^4$, and subject to
the constraint that the row- and column sums in Table~\ref{tab:ProbT2} equal
to the six observable transition probabilities. Two upper bounds are given
by
\begin{align}
\begin{array}{l}
p_{00,01} + p_{00,11} + p_{01,01} + p_{01,11} \leq 1 - \Pr \left( Y_1 = 1 %
\big| D=1\right) , \\[4pt]
p_{00,01} + p_{00,11} + p_{01,01} + p_{01,11} \leq 1 - \Pr \left( Y=(0,0) %
\big| D=0\right) .%
\end{array}
\label{InequT2e2}
\end{align}
Here, the first upper bound follows from the row-sum conditions in Table~\ref%
{tab:ProbT2}, which require that $p_{00,01} + p_{00,11} + p_{01,01} +
p_{01,11} $ is smaller than $\Pr \left( Y=(0,0) \big| D=1\right) + \Pr
\left( Y=(0,1) \big| D=1\right)$, which equals $1 - \Pr \left( Y_1 = 1 \big| %
D=1\right) $. Analogously, the second upper bound follows from the
column-sum conditions in Table~\ref{tab:ProbT2}, which require that $%
p_{00,01} + p_{00,11} + p_{01,01} + p_{01,11} $ is smaller than $\Pr \left(
Y=(0,1) \big| D=0\right) + \Pr \left( Y_1 = 1 \big| D=0\right)$, which
equals $1 - \Pr \left( Y=(0,0) \big| D=0\right)$. Note that if the
destination state is absorbing, the second upper bound in (\ref{InequT2e2})
is $\Pr(Y=(0,1)|D=0)$ which is smaller than for the non-absorbing case.

Regarding the lower bound, notice that $p_{00,01} + p_{01,01}$ cannot be
arbitrarily small, because when shifting probability mass within the column
of Table~\ref{tab:ProbT2} that corresponds to $\Pr \left( Y=(0,1) \big| %
D=0\right)$, we cannot increase the other elements in this column (i.e. $%
p_{10,01} + p_{11,01}$) to more than $\Pr \left( Y_1 = 1 \big| D=1\right)$,
since we would otherwise violate the corresponding row-constraint. By that
argument we find the bound $p_{00,01} + p_{01,01} \geq \Pr \left( Y=(0,1) %
\big| D=0\right) - \Pr \left( Y_1 = 1 \big| D=1\right)$. Together with the
positivity condition on all probabilities we thus obtain
\begin{align}
p_{00,01} + p_{00,11} + p_{01,01} + p_{01,11} &\geq \max \bigg \{ 0 , \, \Pr
\left( Y=(0,1) \big| D=0\right) - \Pr \left( Y_1 = 1 \big| D=1\right) \bigg
\} .  \label{InequT2e3}
\end{align}
The lower bound is the same if the destination state is absorbing. Combining %
\eqref{DefATETSforT2}, \eqref{InequT2e1}, \eqref{InequT2e2} and %
\eqref{InequT2e3}, gives the bounds on $\mathrm{ATETS}_{2}$ summarized in
the following theorem.\footnote{%
Combining \eqref{InequT2e1}, \eqref{InequT2e2} and \eqref{InequT2e3} and $%
\Pr \left( Y_1 = 1 \big| D=1\right) = 1 - \Pr \left( Y_1 = 0 \big| %
D=1\right) $ we obtain
\begin{align*}
\max \bigg \{ 0 , \, \frac{ \Pr \left( Y=(0,1) \big| D=0\right) - 1} { \Pr
\left( Y_1 =0 \big| D=1 \right) } + 1 \bigg \} \leq \mathbb{E} \left(Y_{2}^0 %
\big| Y_{1}^{1}=0 \right) \leq \min \left \{1 , \; \frac{ 1 - \Pr \left(
Y=(0,0) \big| D=0\right) } { \Pr \left( Y_1 =0 \big| D=1 \right) } \right \}
.
\end{align*}
Also using $\eqref{DefATETSforT2}$ and $\mathbb{E}(Y_{2}^{1}|Y_{1}^{1}=0) =
\Pr \left( Y_{2} =1 \big|  Y_{1}=0, D=1 \right)$, and rewriting $\Pr \left(
Y=(0,1) \big| D=0\right)$ and $\Pr \left( Y=(0,0) \big| D=0\right) $ as
products of one-step ahead conditional transition probabilities gives the
bounds in Theorem~\ref{ATETSBasBounds-t} for the case $t=2$.} We find it
convenient to present the theorem for the case of $\mathrm{ATETS}_t$ for
arbitrary $t$. For this we introduce the notation $\overline{Y}%
_{t-1}=(Y_{1},\ldots ,Y_{t-1})$, and we write $0 $ for the vector of zeros.

\begin{theorem}[Bounds on ATETS]
\label{ATETSBasBounds-t} Suppose that Assumption \ref{AssRand3} holds. Let $%
t \in \{2,3,4,\ldots \}$. If $\Pr \left( \overline{Y}_{t-1}=0\,|\,D=1\right)
=0$, then $\mathrm{ATETS}_{t}$ is not defined. If $\Pr \left( \overline{Y}%
_{t-1}=0\,|\,D=1\right) >0$, and also $\Pr( D =1 ) >0$ and $\Pr( D =0 ) >0$,
then we have the bounds
\begin{equation*}
\mathrm{LB}_{t}\leq \mathrm{ATETS}_{t}\leq \mathrm{UB}_{t},
\end{equation*}%
where
\begin{align*}
\mathrm{LB}_{t}& \equiv \Pr (Y_{t}=1\,|\, \overline{Y}_{t-1}=0,D=1) \\
& \qquad -\min \left \{ 1,\frac{1-[1-\Pr (Y_{t}=1\,|\, \overline{Y}%
_{t-1}=0,D=0)]\Pr \left( \overline{Y}_{t-1}=0\,|\,D=0\right) }{\Pr (%
\overline{Y}_{t-1}=0\,|\,D=1)}\right \} , \\
\mathrm{UB}_{t}& \equiv \Pr (Y_{t}=1\,|\, \overline{Y}_{t-1}=0,D=1) \\
& \qquad -\max \left \{ 0,\frac{\Pr (Y_{t}=1\,|\, \overline{Y}%
_{t-1}=0,D=0)\Pr \left( \overline{Y}_{t-1}=0\,|\,D=0\right) -1}{\Pr (%
\overline{Y}_{t-1}=0\,|\,D=1)}+1\right \} .
\end{align*}
\end{theorem}

\noindent \textbf{Proof} See Appendix A.

Notice that the bounds in Theorem~\ref{ATETSBasBounds-t} require no
assumptions beyond random assignment. They allow, for instance, for
arbitrary heterogeneity in treatment response. The bounds exist as long as $%
\Pr \left( \overline{Y}_{t-1}=0\,|\,D=1\right) >0$, because if this
probability is zero, then the subpopulation for which $\mathrm{ATETS}_{t}$
is defined has no members.\footnote{%
The bounds in Theorem~\ref{ATETSBasBounds-t} also involve conditioning on
the event $\overline{Y}_{t-1}=0$ and $D=0$, but we do not need to impose $%
\Pr \left( \overline{Y}_{t-1}=0\,|\,D=0\right) >0$, because all expressions
involving that conditioning set can be rewritten, for example, we have $%
[1-\Pr (Y_{t}=1\,|\, \overline{Y}_{t-1}=0,D=0)]\Pr \left( \overline{Y}%
_{t-1}=0\,|\,D=0\right) = \Pr (Y_{t}=0 \; \& \; \overline{Y}_{t-1}=0 \, |\,
D=0)$.} The conditions $\Pr( D =1 ) >0$ and $\Pr( D =0 ) >0$ guarantee that
both treated and untreated individuals are observed, which is an obvious
condition for any treatment effect estimation.

Next, consider the intuition behind these bounds using the job-bonus
experiment as an illustration. Both the upper and the lower bound are
increasing in the observed transition probability from unemployment to
employment in the treatment group in period $t,$ $\Pr (Y_{t}=1\,|\,
\overline{Y}_{t-1}=0,D=1)$. This follows directly from the fact that we
consider the average effect for treated individuals that remain in
unemployment until time $t$. The bounds also depend on the observed
transition probability in the control group, $\Pr (Y_{t}=1\,|\, \overline{Y}%
_{t-1}=0,D=0)$, but this relationship is more complicated than the
relationship between the bounds and $\Pr (Y_{t}=1\,|\, \overline{Y}%
_{t-1}=0,D=1)$. In general we have that both the upper and the lower bound
are decreasing in $\Pr (Y_{t}=1\,|\, \overline{Y}_{t-1}=0,D=0)$. The reason
for this is that a high transition rate among the unemployed individuals in
the control group allows for a larger counterfactual outcome under no
treatment. Another important determinant of the bounds is the fraction in
the treatment group that remains in unemployment until time $t$, $\Pr (%
\overline{Y}_{t-1}=0\,|\,D=1)$. If this survival probability is small, there
is more selection in the group of treated that remains in unemployment, i.e.
more pronounced dynamic selection, leading to a larger difference between
the upper and the lower bound.

From Theorem \ref{ATETSBasBounds-t} we also have several other implications.
Corollary \ref{Point} shows that if the survival rates under treatment and
control both equal one, i.e., if $\Pr (\overline{Y}_{t-1}=0\,|\,D=0)=1$ and $%
\Pr (\overline{Y}_{t-1}=0\,|\,D=1)=1$, then the dynamic treatment effect $%
\mathrm{ATETS}_{t}$ is point identified. If everyone survives the first $t-1$
periods we have under random treatment assignment in period 1 two groups of
equal composition even in period $t$.

\begin{corollary}[Point identification]
\label{Point} $\mathrm{ATETS}_{t}$ is point identified if both $%
\Pr (\overline{Y}_{t-1}=0\,|\,D=0)=1$ and $\Pr (\overline{Y}%
_{t-1}=0\,|\,D=1)=1$.
\end{corollary}

The information in the bounds depends on the width of the implied interval.
The best case is that the restrictions imposed by the $\max $ and $\min $ in $\mathrm{LB}_{t}$ and $\mathrm{UB}_{t}$ above
are non-binding, and the width of the bounds then becomes
\begin{equation*}
\mathrm{UB}_{t}-\mathrm{LB}_{t}=\frac{2-\Pr (\overline{Y}_{t-1}=0\,|\,D=1)-%
\Pr (\overline{Y}_{t-1}=0\,|\,D=0)}{\Pr (\overline{Y}_{t-1}=0\,|\,D=1)}.
\end{equation*}%
This expression shows that the width of the bound is decreasing in $\Pr (%
\overline{Y}_{t-1}=0\,|\,D=1)$ and $\Pr (\overline{Y}_{t-1}=0\,|\,D=0)$. In
the job-bonus application this implies that the width of the bound is
directly related to the probability that unemployed individuals in the
treatment group and in the control group remain in unemployment until time $%
t $.

\subsection{Arbitrary time to treatment}

\label{s:arbitrary}

So far we have considered the case with treatment assignment at the
beginning of the first period. We now consider a more general case in which
the treatment can start in any time period. We assume that any treated unit
remains treated in the subsequent periods, that is, we assume that treatment
is an absorbing state. Let the potential outcome $Y_{t}^{k}$ be an indicator
of a transition in period $t$ if the treatment started in period $k \le t $,
and since treatment is assumed to be an absorbing state this means that the
unit is treated in all subsequent periods. The potential outcome if
non-treated is denoted by $Y_{t}^{0}$.\footnote{%
Under the no-anticipation assumption in \citeasnoun{Abbring2003}, $%
Y_{t}^{0}$ corresponds to the potential outcome if never-treated, since
no-anticipation assures that the non-treated potential outcome at $t$ equals
the potential outcome at $t$ if never-treated. Without the no-anticipation
assumption the potential outcome, $Y_{t}^{0}$, corresponds to the potential
outcome if non-treated up until $t$, including any anticipatory responses to
information about treatments after $t$.}

Let $D_{t}$ be the indicator of treatment in period $t$ so that a unit with $%
D_{t}=1$ could either be treated or non-treated before $t$. We use the
notation $\overline{D}_{t-1}=(D_{1},\ldots ,D_{t-1})$ and write $1$ and $0$
for the vector of ones and zeros. Note that because the treatment state is
absorbing we have that $\overline D_{t-1}=1 \Leftrightarrow \overline D_t =1$.

With treatment assignments in all periods we need a different randomization
assumption. The relevant random assignment assumption is

\begin{assumption}[Sequential randomization among survivors]
\label{AssRand3_b} For all $t$,
\begin{equation*}
D_{t}\, \bot \, \left \{ Y_{s}^{k} \, :\, k,s=t,t+1,t+2,\ldots \right \} \; %
\bigg|\; \overline{D}_{t-1}=0,\;Y_{t-1}^{0}=\cdots =Y_{1}^{0}=0.
\end{equation*}
\end{assumption}

This assumption implies that treatment is assigned randomly among survivors
that have not been treated before.\footnote{%
Sequential randomization occurs in medical studies in the  Sequential Multiple Assignment Randomized Trial 
(SMART) design, see \citeasnoun{murphy2009screening}
and \citeasnoun{murphy2005experimental}.
}

The treatment effect at $t$ of a treatment started in $k \le t$ is\footnote{%
Note that $\mathrm{ATETS}_{t}(k)$ with $k=1$ is identical to $%
\mathrm{ATETS}_{t}$ considered above. Here, we use the more general
notation, $\mathrm{ATETS}_{t}(k)$, to define the average effects when
treatment could start in any period.}
\begin{equation}  \label{k-effect}
\mathrm{ATETS}_{t}(k)=\mathbb{E}\left( Y_{t}^{k}\, \Big|\, \overline{Y}%
_{t-1}^{k}=0, \overline D_{k-1}=0 \right) -\mathbb{E}\left( Y_{t}^{0}\, \Big|%
\, \overline{Y}_{t-1}^{k}=0, \overline D_{k-1}=0\right) .
\end{equation}
where we average over the subpopulation that started treatment in $k$ and
was not treated before $k$. For the instantaneous treatment effect (if there
is no anticipation effect, the outcome in $k-1$ and earlier is the
non-treated outcome)
\begin{equation*}
\mathrm{ATETS}_{k}(k)=\mathbb{E}\left( Y_{k}^{k}\, \Big|\, \overline{Y}%
_{k-1}^{0}=0, \overline D_{k-1}=0 \right) -\mathbb{E}\left( Y_{k}^{0}\, \Big|%
\, \overline{Y}_{k-1}^{0}=0, \overline D_{k-1}=0\right)
\end{equation*}
Under sequential randomization as in Assumption \ref{AssRand3_b} we have
\begin{equation*}
\mathrm{ATETS}_{k}(k)=\mathbb{E}\left( Y_{k}^{k}\, \Big|\, \overline{Y}%
_{k-1}^{0}=0, D_k=1, \overline D_{k-1}=0 \right) -\mathbb{E}\left(
Y_{k}^{0}\, \Big|\, \overline{Y}_{k-1}^{0}=0, D_k=0 , \overline
D_{k-1}=0\right) =
\end{equation*}
\begin{equation*}
\mathbb{E}\left( Y_{k}\, \Big|\, \overline{Y}_{k-1}=0, D_k=1, \overline
D_{k-1}=0 \right) -\mathbb{E}\left( Y_{k} \, \Big|\, \overline{Y}_{k-1}=0,
D_k=0, \overline D_{k-1}=0\right)
\end{equation*}
so that the instantaneous effect of a treatment starting at $k$ is point
identified.

For the $\mathrm{ATETS}_t(k)$ in \ref{k-effect} we derive the bounds as in
Theorem 1 with $k$ the first period and time of randomization, i.e. in the
role of period 1, and at time $k$ we consider the observations with $%
\overline Y^0_{k-1}=0, \overline D_{k-1}=0$, i.e. the survivors if not
treated that did not receive treatment before $k$ which is the same as $%
\overline Y_{k-1}=0, \overline D_{k-1}=0$. For this subpopulation the data
that enter the bounds are the transition probabilities in $t$ given
treatment starting in $k$ (and continuing until $t$) and given being
assigned to the control group at $k$ (and remaining in the control group
until $t$). The bounds of Theorem 1 apply directly with obvious changes in
the conditioning sets of the probabilities ( condition on $\overline
Y_{k-1}=0, \overline D_{k-1}=0$ in addition to the conditioning variables in
the bounds of Theorem 1).

\section{Bounds on treatment effects on transitions under additional
assumptions}

\label{s:additional}

The bounds in the previous section did not impose any assumptions beyond
random assignment. In this section, we explore the identifying power of
additional assumptions. For sake of presentation we will focus on
identification of $\mathrm{ATETS}_{t}$.\footnote{%
Assumptions that tighten $\mathrm{ATETS}_{t}(k)$ with $k>1$ follow using
similar reasoning.} The assumptions that we make
are implicit in parametric models such as the MPH model, and also in the discrete duration models and
structural models presented in this section.

As a background consider the following discrete duration model for the
control and treated outcomes, for individual $i$ in period $t$,
\begin{eqnarray}
Y_{it}^{0} &=&I(\alpha _{t}+V_{i}-\varepsilon _{it}^{0}\geq 0),  \notag \\
Y_{it}^{1} &=&I(\alpha _{t}+\gamma _{it}+V_{i}-\varepsilon _{it}^{1}\geq 0).
\label{eqDiscr}
\end{eqnarray}%
This discrete duration model has a composite error that is the sum of
unobserved heterogeneity $V_{i}$ and a random shock $\varepsilon _{it}$. {%
Here, $\alpha _{t}$ is a time specific effect, and $\gamma _{it}$ drives the
systematic differences between treated and non-treated outcomes. } The model
allows for different random shocks under control, $\varepsilon _{it}^{0}$
and treatment, $\varepsilon _{it}^{1}$. These random shocks are assumed to
be independent, but even in this case the potential outcomes are positively
correlated through their dependence on $V_{i}$. A more traditional model has
the same random shock under control and treatment, $\varepsilon _{it}$, but
this is a more restrictive model. In the sequel we start from the more
general model in (\ref{eqDiscr}) to illustrate the additional assumptions.

\subsection{Monotone Treatment Response}

The first assumption is Monotone Treatment Response (MTR). The assumption is
that the effect is either positive or negative for all units in all periods.
In terms of the discrete duration model example in (\ref{eqDiscr}), the
assumption is that $\gamma _{it}\leq 0$ for all $i,t$ or $\gamma _{it}\geq 0$
for all $i,t$. That is, we do not assume a specific direction of the effect,
merely that the effect goes in the same direction for all units. For the
job-bonus experiment considered in this paper this assumption rules out that
the bonus offer increases the transition rate for some unemployed
individuals and decreases the transition rate for others. The assumption is
similar to the MTR assumption introduced by \citeasnoun{Manski1997} and %
\citeasnoun{Manski2000}.

To formally define MTR in our framework we denote the event of survival
under treatment and no-treatment by $S_{t}$, that is, $S_{t}$ is the event
that $\overline{Y}_{t}^{1}=0$ and $\overline{Y}_{t}^{0}=0 $. We have

\begin{assumption}[Monotone Treatment Response (MTR)]
\label{MTR-gen} Either
\begin{equation*}
\Pr \left( \left. Y_{t}^{1}=1\right \vert S_{t-1},\,V\right) \geq \Pr \left(
\left. Y_{t}^{0}=1\right \vert S_{t-1},\,V\right) ,
\end{equation*}%
for all $t$, or
\begin{equation*}
\Pr \left( \left. Y_{t}^{1}=1\right \vert S_{t-1},\,V\right) \leq \Pr \left(
\left. Y_{t}^{0}=1\right \vert S_{t-1},\,V\right) ,
\end{equation*}%
for all $t$. Here, $V$ can be any known or unknown vector of  individual
specific characteristics (both observed and unobserved) that are constant over time.\footnote{%
In particular, $V$ could identify the individual $i$ uniquely. In that case
the assumption simply becomes that we have either $\Pr \left( \left.
Y_{it}^{1}=1\right \vert S_{i,t-1}\right) \geq \Pr \left( \left.
Y_{it}^{0}=1\right \vert S_{i,t-1}\right) $, for all $i,t$, or $\Pr \left(
\left. Y_{it}^{1}=1\right \vert S_{i,t-1}\right) \leq \Pr \left( \left.
Y_{it}^{0}=1\right \vert S_{i,t-1}\right) $, for all $i,t$. This was the
formulation of the assumption used in a previous version of this paper.}
\end{assumption}

This assumption can be relaxed at the expense of more complicated bounds.
The assumption is that the effect goes in the same direction for all units.
This is consistent with a discrete duration model that allows the random and
independent shocks $\varepsilon _{it}^{1}$ and $\varepsilon _{it}^{0}$ to
differ, but restricts the sign of $\gamma _{it}$.

\subsection{Common Shocks}

The next assumption restricts the joint distribution of potential outcomes
in the treatment arms. The assumption essentially imposes that the outcomes
in both treatment arms involve the same random shock. In terms of the
discrete duration model example in (\ref{eqDiscr}), the assumption is that $%
\varepsilon _{it}^{1}=\varepsilon _{it}^{0}=\varepsilon _{it}$, so that the random shock $\varepsilon _{it}$ is the same for both  treatment states.
 Thus, if $\gamma _{it}\leq 0$ then the treated have a larger
survival probability in $t$. Therefore the event that $i$ survives in $t$ if
not treated, i.e. $Y_{it}^{0}=0$, is equivalent to $\varepsilon _{it}\geq
\alpha _{t}+V_{i}$, so that this event implies that $\varepsilon _{it}\geq
\alpha _{t}+\gamma _{it}+V_{i}\geq 0$, i.e. $Y_{it}^{1}=0$. In a structural model the random shocks often satisfy this restrictions, as is illustrated in a simple job search model below.

The formal statement of the assumption is as follows.

\begin{assumption}[Common Shocks (CS)]
\label{Single} For all $t$
\begin{equation*}
\Pr (Y_{t}^{1}=0|S_{t-1},V)\geq \Pr (Y_{t}^{0}=0|S_{t-1},V)\quad \Rightarrow
\quad \Pr (Y_{t}^{1}=0|S_{t-1},Y_{t}^{0}=0,V)=1,
\end{equation*}%
\begin{equation*}
\Pr (Y_{t}^{1}=0|S_{t-1},V)\leq \Pr (Y_{t}^{0}=0|S_{t-1},V)\quad \Rightarrow
\quad \Pr (Y_{t}^{0}=0|S_{t-1},Y_{t}^{1}=0,V)=1.
\end{equation*}%
Here, again, $V$ can be any known or unknown vector of   individual specific
characteristics (both observed and unobserved) that are constant over time.\footnote{%
In particular, $V$ could identify the individual $i$ uniquely. In that case
the assumption simply becomes that for $i$ we have $\Pr
(Y_{it}^{1}=0|S_{i,t-1})\geq \Pr (Y_{it}^{0}=0|S_{i,t-1})\quad \Rightarrow
\quad \Pr (Y_{it}^{1}=0|S_{i,t-1},Y_{it}^{0}=0)=1,$ and $\Pr
(Y_{it}^{1}=0|S_{i,t-1})\leq \Pr (Y_{it}^{0}=0|S_{i,t-1})\quad \Rightarrow
\quad \Pr (Y_{it}^{0}=0|S_{i,t-1},Y_{it}^{1}=0)=1$.}
\end{assumption}

In the job-bonus application the intuition behind this assumption is that CS
implies that all random events leading to a job offer and employment are the
same irrespective if a specific unemployed individual is randomized to the
treatment group or to the control group.

Assumption \ref{Single} is satisfied in many standard structural models.
Consider for instance a non-stationary job search model for an unemployed
individual as in \citeasnoun{VandenBerg1990} or \citeasnoun{Meyer1996}. The
treatment is a re-employment bonus as discussed in Section 5 below. In each
period a job offer is obtained with probability $p(t,V_{i})$. Let $Y_{of,it}$
be the indicator of an offer in period $t$ and $Y_{of,it}=I(\varepsilon
_{of,it}\in A(t,V_{i}))$ with $A(t,V_{i})$ a set. If the job offer is not
under control of $i$, the arrival process is the same under treatment and
control. The reservation wage is denoted by $\xi _{it}^{1}$ for the treated
and $\xi _{it}^{0}$ for the controls. In general (see \citeasnoun{Meyer1996}%
) $\xi ^{1}(t,V_{i})\leq \xi ^{0}(t,V_{i})$, so that if $H$ is the wage
offer c.d.f. we have the acceptance probabilities $1-H(\xi
^{1}(t,V_{i}))\geq 1-H(\xi ^{0}(t,V_{i}))$. The acceptance indicators are $%
Y_{ac,it}^{0}=I(\varepsilon _{w,it}\geq \xi ^{0}(t,V_{i}))$ and $%
Y_{ac,it}^{1}=I(\varepsilon _{w,it}\geq \xi ^{1}(t,V_{i}))$ with $%
\varepsilon _{w,it}$ the wage offer. Because $%
Y_{it}^{0}=Y_{of,it}Y_{ac,it}^{0}$ and $Y_{it}^{1}=Y_{of,it}Y_{ac,it}^{1}$,
we have
\begin{equation*}
Y_{it}^{1}=0\quad \Rightarrow \quad Y_{it}^{0}=0,
\end{equation*}%
so that Assumption \ref{Single} is satisfied.

\subsection{Positively correlated outcomes}

The third assumption concerns the relation between the counterfactual
outcomes over time. Let us introduce the assumption for the two period case. If we compare the transition probability $\Pr
(Y_{2}^{0}=1|Y_{1}^{1}=0,Y_{1}^{0}=0)$ to $\Pr
(Y_{2}^{0}=1|Y_{1}^{1}=1,Y_{1}^{0}=0)$, i.e. the probability of a transition
in period 2 if no treatment was received in periods 1 and 2 given survival
with or without treatment in period 1 to the same probability given survival
without but not with treatment in period 1, then it is reasonable to assume
that the former probability is not larger than the latter. Individuals with $%
Y_{1}^{1}=0,Y_{1}^{0}=0$ have characteristics that make them not leave the
initial state as opposed to individuals with $Y_{1}^{1}=1,Y_{1}^{0}=0$ that
have characteristics that make them leave the initial state if treated in
period 1. If the variables that affect the transition out of the initial
state are positively correlated between periods, then
\begin{equation}
\Pr (Y_{2}^{0}=1|Y_{1}^{1}=0,Y_{1}^{0}=0)\leq \Pr
(Y_{2}^{0}=1|Y_{1}^{1}=1,Y_{1}^{0}=0)\text{.}  \label{eqPCO1}
\end{equation}

As before we motivate the assumption using the discrete duration model in (%
\ref{eqDiscr}). Consider (\ref{eqPCO1}). By the discrete duration model the
conditioning events are if no transition (i.e., if $Y_{1}^{1}=0,Y_{1}^{0}=0$%
)
\begin{equation*}
V_{i}-\varepsilon _{i1}^{0}<-\alpha _{1},V_{i}-\varepsilon _{i1}^{1}<-\alpha
_{1}-\gamma _{i1},
\end{equation*}%
and if a transition in 1 if treated (i.e., if $Y_{1}^{1}=1,Y_{1}^{0}=0$)
\begin{equation*}
V_{i}-\varepsilon _{i1}^{0}<-\alpha _{1},V_{i}-\varepsilon _{i1}^{1}\geq
-\alpha _{1}-\gamma _{i1}.
\end{equation*}

Thus, if $V_{i}-\varepsilon _{i1}^{0}$ is positively correlated with $%
V_{i}-\varepsilon _{i2}^{0}$, then (\ref{eqPCO1}) holds, since then $\Pr
(Y_{2}^{0}=1)$ is at least as large for the subpopulation with $%
Y_{1}^{1}=0,Y_{1}^{0}=0$ as for the subpopulation with $%
Y_{1}^{1}=1,Y_{1}^{0}=0$. We call this positively correlated outcomes. An
analogous argument can be made for the relation between $\Pr
(Y_{2}^{0}=1|Y_{1}^{1}=0,Y_{1}^{0}=1)$ and $\Pr
(Y_{2}^{0}=1|Y_{1}^{1}=0,Y_{1}^{0}=0)$, as well as for $\Pr (Y_{2}^{1}=1)$\
for different subpopulations.

Formally, for arbitrary $t$ we have

\begin{assumption}[Positively Correlated Outcomes (PCO)]
\label{PCO} For all $m=1,\ldots ,t-1$
\begin{eqnarray*}
\Pr (Y_{t}^{0}=1|Y_{m}^{1}=1,\overline{Y}_{m-1}^{1}=0,\overline{Y}%
_{t-1}^{0}=0) &\geq &\Pr (Y_{t}^{0}=1|\overline{Y}_{t-1}^{1}=0,\overline{Y}%
_{t-1}^{0}=0), \\
\Pr (Y_{t}^{1}=1|Y_{m}^{1}=1,\overline{Y}_{m-1}^{1}=0,\overline{Y}%
_{t-1}^{0}=0) &\geq &\Pr (Y_{t}^{1}=1|\overline{Y}_{t-1}^{1}=0,\overline{Y}%
_{t-1}^{0}=0), \\
\Pr (Y_{t}^{0}=1|\overline{Y}_{t-1}^{1}=0,Y_{m}^{0}=1,\overline{Y}%
_{m-1}^{0}=0) &\geq &\Pr (Y_{t}^{0}=1|\overline{Y}_{t-1}^{1}=0,\overline{Y}%
_{t-1}^{0}=0), \\
\Pr (Y_{t}^{1}=1|\overline{Y}_{t-1}^{1}=0,Y_{m}^{0}=1,\overline{Y}%
_{m-1}^{0}=0) &\geq &\Pr (Y_{t}^{1}=1|\overline{Y}_{t-1}^{1}=0,\overline{Y}%
_{t-1}^{0}=0).
\end{eqnarray*}
\end{assumption}

For the job-bonus application PCO\ has several implications. As an
illustration, consider two groups consisting of unemployed who find and
unemployed who do not find employment in the first period if non-treated. In
this case PCO implies that in the second period, the transition rate under
treatment on average is weakly larger in the former group compared to the
latter. This holds if the ranking of the unemployed individuals in terms of
the characteristics that determine job offers, such as ability, experience
and job search effort, remains the same during the entire unemployment spell.

Note that the motivating example above shows that PCO does not imply nor is
implied by MTR or CS. The CS assumption is on the contemporaneous
correlation of random shocks while PCO relates to a (positive) relation of
the combined random error over time. Since the latter in general contains an
important individual effect, positive correlation is not a strong assumption.

\subsection{Bounds under the additional assumptions}

We now obtain bounds on $\mathrm{ATETS}$ for arbitrary $t$ when we compare a
treatment started in period 1 to no treatment in all periods. Bounds under
MTR and CS are given in Theorem \ref{ATETSBoundsMTR-t} and Theorem \ref%
{ATETS-Th-PCO-2} provides bounds under PCO. Bounds under all three
additional assumptions are in Theorem \ref{ATETSBoundsMTRPCO-t}.

\begin{theorem}[Bounds on ATETS under MTR and CS for $t$ periods]
\label{ATETSBoundsMTR-t} Let the Assumptions \ref{AssRand3}, \ref{MTR-gen},
and \ref{Single} hold. Let $t\in \{2,3,4,\ldots \}$. If $\Pr \left(
\overline{Y}_{t-1}=0|D=1\right) =0$, then $\mathrm{ATETS}_{t}$ is not
defined.

If $\Pr \left( \overline{Y}_{t-1}=0|D=1\right) >0$, and also $\Pr (D=1)>0$
and $\Pr (D=0)>0$, then we have the bounds
\begin{equation*}
\mathrm{LB}_{t}\leq \mathrm{ATETS}_{t}\leq \mathrm{UB}_{t},
\end{equation*}%
where
\begin{align*}
\mathrm{LB}_{t}& =\Pr (Y_{t}=1|\overline{Y}_{t-1}=0,D=1) \\
& \quad -\min \Bigg \{1,\;1+\frac{\Pr (Y_{t}=1|\overline{Y}_{t-1}=0,D=0)\Pr
\left( \overline{Y}_{t-1}=0|D=0\right) }{\Pr (\overline{Y}_{t-1}=0|D=1)} \\
& \qquad \qquad \qquad \quad -\frac{\min \left \{ \Pr (\overline{Y}%
_{t-1}=0|D=1),\Pr (\overline{Y}_{t-1}=0|D=0)\right \} }{\Pr (\overline{Y}%
_{t-1}=0|D=1)}\Bigg \}, \\
\mathrm{UB}_{t}& =\Pr (Y_{t}=1|\overline{Y}_{t-1}=0,D=1) \\
& \quad -\max \Bigg \{0\;,\frac{[\Pr (Y_{t}=1|\overline{Y}%
_{t-1}=0,D=0)-1]\Pr \left( \overline{Y}_{t-1}=0|D=0\right) }{\Pr (\overline{Y%
}_{t-1}=0|D=1)} \\
& \quad \qquad \qquad \qquad +\frac{\min \left \{ \Pr (\overline{Y}%
_{t-1}=0|D=1),\Pr (\overline{Y}_{t-1}=0|D=0)\right \} }{\Pr (\overline{Y}%
_{t-1}=0|D=1)}\Bigg \}.
\end{align*}
\end{theorem}

\noindent \textbf{Proof} See Appendix A. \newline

Assumption \ref{MTR-gen} states that the treatment effect is either
non-negative or non-positive for all i. Since in period 1 we can estimate
the ATETS directly because there is no dynamic selection yet, the
possibility that MTR holds with a non-positive effect, can be excluded if
the ATETS in period 1 is non-negative. If we make the stronger assumption
that the effect has the same sign for all i and for all t then a
non-negative ATETS in period 1 excludes non-positive MTR in all periods. In
that case the ATETS is non-negative in all time periods and this improves
the lower bound on the ATETS, but has no effect on the upper bound that is
between 0 and 1. The lower bound on the ATETS if non-negative MTR holds is%
\footnote{%
In the same way, if the ATETS in period 1 is non-positive, the possibility
that MTR holds with a non-negative effect can be excluded,affecting the
upper bound in an obvious way.}:%
\begin{eqnarray*}
\mathrm{LB}_{t} &=&\max \Bigg \{0,\Pr (Y_{t}=1|\overline{Y}_{t-1}=0,D=1) \\
&&-\frac{\Pr (Y_{t}=1|\overline{Y}_{t-1}=0,D=0)\Pr \left( \overline{Y}%
_{t-1}=0|D=0\right) }{\Pr (\overline{Y}_{t-1}=0|D=1)}\Bigg \}.
\end{eqnarray*}%
If MTR can change sign between periods we would require prior knowledge of
the sign in each time period to improve on the bounds in Theorem 2.

\begin{theorem}[Bounds on ATETS under PCO for $t$ periods]
\label{ATETS-Th-PCO-2} Let Assumptions \ref{AssRand3} and \ref{PCO} hold.
Let $t\in \{2,3,4,\ldots \}$. If $\Pr \left( \overline{Y}_{t-1}=0|D=1\right)
=0$, then $\mathrm{ATETS}_{t}$ is not defined.

If $\Pr \left( \overline{Y}_{t-1}=0|D=1\right) >0$ and $\Pr (Y_{s}=0|%
\overline{Y}_{s-1}=0,D=1)+\Pr (Y_{s}=0|\overline{Y}_{s-1}=0,D=0)-1>0$ for
all $s=1,\ldots ,t-1$, and also $\Pr (D=1)>0$ and $\Pr (D=0)>0$, then we
have the bounds
\begin{equation*}
\mathrm{LB}_{t}\leq \mathrm{ATETS}_{t}\leq \mathrm{UB}_{t},
\end{equation*}%
where
\begin{align*}
& \mathrm{LB}_{t}=\Pr (Y_{t}=1|D=1,\overline{Y}_{t-1}=0)-1+\frac{1-\Pr
(Y_{t}=1|\overline{Y}_{t-1}=0,D=0)}{\Pr (\overline{Y}_{t-1}=0|D=1)} \\
& \qquad \qquad \times \prod_{s=1}^{t-1}[\Pr (Y_{s}=0|\overline{Y}%
_{s-1}=0,D=1)+\Pr (Y_{s}=0|\overline{Y}_{s-1}=0,D=0)-1], \\
& \mathrm{UB}_{t}=\Pr (Y_{t}=1|D=1,\overline{Y}_{t-1}=0) \\
& -\max \left \{ 0,\frac{(\Pr (Y_{t}=1|\overline{Y}_{t-1}=0,D=0)-1)\Pr
\left( \overline{Y}_{t-1}=0|D=0\right) }{\prod_{s=1}^{t-1}[\Pr (Y_{s}=0|%
\overline{Y}_{s-1}=0,D=1)+\Pr (Y_{s}=0|\overline{Y}_{s-1}=0,D=0)-1]}+1\right
\} .
\end{align*}%
If $\Pr \left( \overline{Y}_{t-1}=0|D=1\right) >0$ and $\Pr (Y_{s}=0|%
\overline{Y}_{s-1}=0,D=1)+\Pr (Y_{s}=0|\overline{Y}_{s-1}=0,D=0)-1\leq 0$
for some $s\leq t$, then we have the bounds
\begin{equation*}
\Pr (Y_{t}=1|\overline{Y}_{t-1}=0,D=1)-1\leq \mathrm{ATETS}_{t}\leq \Pr
(Y_{t}=1|\overline{Y}_{t-1}=0,D=1).
\end{equation*}
\end{theorem}

\noindent \textbf{Proof} See Appendix A.

\begin{theorem}[Bounds on ATETS under MTR, CS and PCO for $t$ periods]
\label{ATETSBoundsMTRPCO-t} Let the Assumptions \ref{AssRand3}, \ref{MTR-gen}%
, \ref{Single}, and \ref{PCO} hold. Let $t\in \{2,3,4,\ldots \}$. If $\Pr
\left( \overline{Y}_{t-1}=0|D=1\right) =0$, then $\mathrm{ATETS}_{t}$ is not
defined.

If $\Pr \left( \overline{Y}_{t-1}=0|D=1\right) >0$, and also $\Pr (D=1)>0$
and $\Pr (D=0)>0$, then we have the following bounds
\begin{equation*}
\mathrm{LB}_{t}\leq \mathrm{ATETS}_{t}\leq \mathrm{UB}_{t},
\end{equation*}%
where
\begin{align*}
\mathrm{LB}_{t}& =\Pr (Y_{t}=1|D=1,\overline{Y}_{t-1}=0)-1+\frac{1-\Pr
(Y_{t}=1|\overline{Y}_{t-1}=0,D=0)}{\Pr (\overline{Y}_{t-1}=0|D=1)} \\
& \qquad \qquad \qquad \qquad \qquad \times \min \left \{ \Pr (\overline{Y}%
_{t-1}=0|D=1),\Pr (\overline{Y}_{t-1}=0|D=0)\right \} , \\
\mathrm{UB}_{t}& =\Pr (Y_{t}=1|D=1,\overline{Y}_{t-1}=0) \\
& \quad -\max \left \{ 0,\frac{(\Pr (Y_{t}=1|\overline{Y}_{t-1}=0,D=0)-1)\Pr
\left( \overline{Y}_{t-1}=0|D=0\right) }{\min \left \{ \Pr (\overline{Y}%
_{t-1}=0|D=1),\Pr (\overline{Y}_{t-1}=0|D=0)\right \} }+1\right \} .
\end{align*}
\end{theorem}

\noindent \textbf{Proof} See Appendix A.


\section{Inference}

\label{s:inference}

Initially, for a given time period $t$, we consider inference on $\theta_0 =
\mathrm{ATETS}_t$ based on the identification result in Theorem~\ref%
{ATETSBasBounds-t}. We assume that $\mathrm{Pr}(\overline Y_{t-1}=0|D=1) > 0$%
. The bounds in the theorem can then be expressed as
\begin{align}
\max(a_1,a_2) =: \ell \leq \theta_0 \leq u := \min(a_3,a_4),  \label{bounds}
\end{align}
with
\begin{align*}
a_1 &= a_3 -1 ,  \notag \\
a_2 &= a_3 - \frac{1-[1 - \mathrm{Pr}(Y_t=1|\overline Y_{t-1}=0,D=0)]
\mathrm{Pr}(\overline Y_{t-1}=0|D=0) } { \mathrm{Pr}(\overline Y_{t-1}=0|D=1)%
} ,  \notag \\
a_3 &= \mathrm{Pr}( Y_t=1 | \overline Y_{t-1} = 0,D=1) ,  \notag \\
a_4 &= a_3 - 1 + \frac{1-\mathrm{Pr}(Y_t=1|\overline Y_{t-1}=0,D=0) \mathrm{%
Pr}(\overline Y_{t-1}=0|D=0)} { \mathrm{Pr}(\overline Y_{t-1}=0|D=1)}.
\end{align*}
If we observe an iid sample $\{ (Y_{i1}, Y_{i2}, \ldots, Y_{it}, D_i ) , i
\in 1,\ldots,n \} $, then the sample analog of $a=(a_1,a_2,a_3,a_4)^{\prime
} $ can easily be constructed, for example
\begin{align*}
\widehat a_3 &= \frac{ \frac 1 n \sum_{i=1}^n \mathbbm{1}( Y_{it}=1 , Y_{i1}
= 0 , Y_{i2} = 0, \ldots, Y_{i,t-1} = 0 , D_i =0) } { \frac 1 n \sum_{i=1}^n %
\mathbbm{1}(Y_{i1} = 0 , Y_{i2} = 0, \ldots, Y_{i,t-1} = 0 , D_i =0) } , &
\widehat a_1 &= \widehat a_3 -1 ,
\end{align*}
and analogously for $\widehat a_2$ and $\widehat a_4$. It is easy to show
that as the sample size $n$ goes to infinity
\begin{align}
\sqrt{n} ( \widehat a - a ) \Rightarrow \mathcal{N}(0, \Sigma_a),
\label{Alimit}
\end{align}
and we can construct a consistent estimator $\widehat \Sigma_a$ of the $4
\times 4$ matrix $\Sigma_a$ (for example, we use bootstrapping to calculate $%
\widehat \Sigma_a$ in our application in Section 6). In the following we
assume that $\Sigma_{a,kk}>0$ for all $k=1,2,3,4$.\footnote{%
Since $\widehat a_1$ and $\widehat a_3$ are perfectly correlated we have $%
\Sigma_a v = 0$ for the vector $v = (1,-1,0,0)^{\prime }$, implying that $%
\mathrm{rank}(\Sigma_a) \leq 3$, but this rank deficiency turns out not to
be important for our purposes.}

The identification results in Theorem~2 for $\theta _{0}=\mathrm{ATETS}_{t}$
can also be expressed as $\max (a_{1},\min (a_{2},a_{3}))\leq \theta
_{0}\leq \min (a_{4},\max (a_{5},a_{6}))$, with appropriate definition of $%
a=(a_{1},a_{2},a_{3},a_{4},a_{5},a_{6})^{\prime }$, whose estimator is again
jointly normally distributed asymptotically, and the inference discussion
below can be easily generalized to this case. Similarly with Theorem~3 and 4.

\subsection{Connection to the Moment Inequality Literature}

The inference problem for $\theta_0$ that is summarized by (\ref{bounds})
and (\ref{Alimit}) is asymptotically equivalent to an inference problem on a
finite number of moment inequalities that is well-studied in the literature,
for example in \citeasnoun{ChernozhukovHongTamer2007}, %
\citeasnoun{RomanoShaikh2008}, \citeasnoun{Rosen2008}, %
\citeasnoun{AndrewsGuggenberger2009}, \citeasnoun{AndrewsSoares2010}, and %
\citeasnoun{AndrewsBarwick2012}. To make this connection explicit we define

\begin{align*}
m(\theta) & := \left(
\begin{array}{c}
\Sigma_{a,11}^{-1/2} ( a_1 - \theta ) \\
\Sigma_{a,22}^{-1/2} ( a_2 - \theta ) \\
\Sigma_{a,33}^{-1/2} ( \theta - a_3 ) \\
\Sigma_{a,44}^{-1/2} ( \theta - a_4 )%
\end{array}
\right) , & \widehat m(\theta) & := \left(
\begin{array}{c}
\widehat \Sigma_{a,11}^{-1/2} ( \widehat a_1 - \theta) \\
\widehat \Sigma_{a,22}^{-1/2} ( \widehat a_2 - \theta) \\
\widehat \Sigma_{a,33}^{-1/2} ( \theta - \widehat a_3 ) \\
\widehat \Sigma_{a,44}^{-1/2} ( \theta - \widehat a_4)%
\end{array}
\right) .
\end{align*}

The bounds (\ref{bounds}) can then equivalently be expressed as $m(\theta_0)
\leq 0$, which is analogous to imposing four moment inequalities.\footnote{$%
m(\theta)$ is not actually a moment function, but has a slightly more
complicated structure (e.g.~$a_3$ is a conditional probability that can be
expressed as the ratio between two moments). This, however, does not matter
for the asymptotic analysis since the estimator $\widehat m(\theta)$ has the
same first order asymptotic properties as it would have in the moment
inequality case. We can therefore fully draw on the insights of the existing
literature.} For convenience we have normalized $m(\theta)$ such that each
component of $\sqrt{n}\left[ \widehat m(\theta) - m(\theta) \right]$ has
asymptotic variance equal to one. Using (\ref{Alimit}) we obtain $\sqrt{n}%
\left[ \widehat m(\theta) - m(\theta) \right] \Rightarrow \mathcal{N}(0 ,
\Sigma_m)$, where $\Sigma_m = A \Sigma_{a} A$, with $A = \mathrm{diag}%
(\Sigma_{a,11}^{-1/2},\Sigma_{a,22}^{-1/2},- \Sigma_{a,33}^{-1/2},-
\Sigma_{a,44}^{-1/2} )$. An estimator $\widehat \Sigma_m$ can be constructed
analogously.

All the papers on moment inequalities cited above start from choosing an
objective function (or criterion function, or test statistics), whose sample
version we denote by $\widehat Q(\theta)$, and then construct a confidence
set for $\theta_0$ as
\begin{align}
\widehat \Theta(C_{1-\alpha}) &= \{ \theta \in \mathbb{R}: n \widehat
Q(\theta) \leq C_{1-\alpha} \} ,
\end{align}
where $C_{1-\alpha} \geq 0$ is a critical value that is chosen such that
confidence $1-\alpha$ is achieved asymptotically, i.e. $\lim_{n \rightarrow
\infty} \mathrm{Pr}( \theta_0 \in \widehat \Theta(C_{1-\alpha}) ) \geq
1-\alpha$.\footnote{%
As discussed in e.g. \citeasnoun{AndrewsSoares2010}, it is important that
the coverage probability is asymptotically bounded by $1-\alpha$ uniformly
over $\theta_0$ and over the distribution of the observables. We have only
formulated the pointwise condition here to keep the presentation simple.}
Various objective functions have been considered in the literature. For
example, the objective function considered in %
\citeasnoun{ChernozhukovHongTamer2007} reads in our notation $\widehat
Q(\theta) = \left \| [ \widehat m(\theta) ]_+ \right \|^2$, where $\|. \|$
refers to the Euclidian norm, and $[\widehat m(\theta)]_+ := \max(0,
\widehat m(\theta))$, applied componentwise to the vector $\widehat
m(\theta) $.

\subsection{Construction of Confidence Intervals}

Our specific inference problem is easier than the general inference problem
for moment inequalities, because in our case the parameter $\theta_0$ is
just a scalar, and the total number of inequalities is relatively small. Our
goal in the following is therefore to outline a concrete method of how to
construct a confidence interval in that special case.

We choose the objective function $\widehat Q(\theta) = \left \| [ \widehat
m(\theta) ]_+ \right \|^2_\infty$, where $\|.\|_\infty$ is the infinity norm,%
\footnote{%
This is special case of the ``test function'' $S_3(m,\Sigma)$ introduced in
equation (3.6) of \citeasnoun{AndrewsSoares2010}, with $p_1=1$ and $v=0$ in
their notation.} i.e.~we have $\widehat Q(\theta) = \max \{ 0, \widehat
m_1(\theta), \widehat m_2(\theta), \widehat m_3(\theta), \widehat
m_4(\theta) \}^2$. This objective function is convenient for our purposes,
because the confidence set defined above then takes the intuitive form
\begin{align}
& \widehat \Theta(C_{1-\alpha})  \notag \\
&= \left[ \max \left( \widehat a_1 - \frac{c_{1-\alpha} \widehat
\Sigma_{a,11}^{1/2}} {\sqrt{n}} , \widehat a_2 - \frac{c_{1-\alpha} \widehat
\Sigma_{a,22}^{1/2} } {\sqrt{n}} \right) , \min \left( \widehat a_3 + \frac{%
c_{1-\alpha} \widehat \Sigma_{a,33}^{1/2} } {\sqrt{n}} , \widehat a_4 +
\frac{c_{1-\alpha} \widehat \Sigma_{a,44}^{1/2} } {\sqrt{n}} \right) \right]
,  \label{SimpleConfidence}
\end{align}
where $c_{1-\alpha} := \sqrt{C_{1-\alpha}}$. This confidence interval can be
constructed very easily.

\subsubsection*{Most Robust Critical Value}

The critical value $c_{1-\alpha}$ still needs to be chosen. The problem with
choosing the critical value in moment inequality problems is that this
choice depends on the unknown slackness vector $m(\theta_0)$, which
indicates whether each inequality $m_k(\theta_0) \leq 0$ is binding, close
to binding, or far from binding. It is known, however, that the largest
(``worst case'') critical value needs to be chosen if $m(\theta_0) = 0$,
i.e.~if all moment inequalities are binding at the true parameter. To find
this critical value one can use the fact that in this worst case $n \widehat
Q(\theta)$ is asymptotically distributed as $\left \| [ Z ]_+
\right
\|_\infty^2$, where $Z \sim \mathcal{N}(0,\Sigma_m)$ is a random
four vector. Using the estimator $\widehat \Sigma_m$ one can simulate this
distribution. However, it can easily be shown that the $1-\alpha$ quantile
of $\left \| [ Z ]_+ \right \|_\infty$ is always smaller or equal to the
following conservative critical value
\begin{align}
c_{1-\alpha} = \Phi^{-1}\left( 1 - \frac{\alpha} 4 \right),
\label{CriticalSimple}
\end{align}
where $\Phi^{-1}$ is the quantile function (the inverse cdf) of the standard
normal distribution. The factor $1/4$ that appears here reflects the fact
that we have four moment inequalities. Combining equations (\ref%
{SimpleConfidence}) and (\ref{CriticalSimple}) provides a confidence
interval that is uniformly valid, i.e.~whose asymptotic size is bounded by $%
\alpha$, independent of what the true values of $a_1$, $a_2$, $a_3$ and $a_4$
are.

\subsubsection*{Critical Value for the Case $\ell \ll u$}

The critical values based on the ``worst case'' where all inequalities are
binding ($m(\theta_0)=0$) can be very conservative if one or multiple
inequalities are far from binding ($m_k(\theta_0) \ll 0$).\footnote{%
In addition, the formula (\ref{CriticalSimple}) only provides an upper bound
for the optimal critical value at $m(\theta_0) = 0$, but this second issue
is often not very severe. For example, for $\alpha=0.05$ and $\Sigma_m =
\mathbb{I}_4$ one finds by simulation that the $0.95$ quantile of $\left \|
[ Z ]_+ \right \|_\infty$, with $Z \sim \mathcal{N}(0,\Sigma_m)$, is $%
c_{0.95} = 2.234$, while the much easier to computer conservative critical
value in (\ref{CriticalSimple}) is $\Phi^{-1}\left( 0.9875 \right) = 2.241$.}
Furthermore, for the inference on $\theta_0 = \mathrm{ATETS}_t$ based on
Theorem~1, with $a$'s as given above, it can easily be shown that if $%
\mathrm{Pr}(\overline Y_{t-1}=0|D=1)>0$ and $\mathrm{Pr}(\overline
Y_{t-1}=0|D=0)<1$, then we have $\max(a_1,a_2) =: \ell < u := \min(a_3,a_4)$%
, implying that $m(\theta_0)=0$ is impossible. However, what matters for the
coverage rate of the confidence interval for a finite sample is not whether $%
\ell < u$, but whether the difference $u - \ell$ is large relative to the
standard deviations $\Sigma_{a,kk}^{1/2}$ of the $\widehat a_k$, $k=1,2,3,4$%
. This is what we mean by $\ell \ll u$ in the subsection title above.

To formalize this one can consider a pretest of the hypothesis $H_0: \ell =
u $, against the alternative $H_a: \ell < u$, with pretest size $\alpha_n^{%
\mathrm{pre}}$ chosen to be very small, e.g.~$\alpha_n^{\mathrm{pre}}=0.001
\ll \alpha$.\footnote{%
Theoretically one can assume $\alpha_n^{\mathrm{pre}} \rightarrow 0$ as $n
\rightarrow \infty$ to avoid asymptotic size distortions due to the pretest.}
If the pretest is not rejected, then the critical value (\ref{CriticalSimple}%
) should be chosen. If the pretest is rejected, then the two problems of
choosing a suitable lower and upper bound for the confidence interval $%
\widehat \Theta$ completely decouple, because with high confidence we know
that for any $\theta$ only one of those bounds can be binding at the same
time, implying that at most two of the moment inequalities $m(\theta_0) \leq
0$ can be binding. In this latter case we can therefore choose the less
conservative critical value
\begin{align}
c_{1-\alpha} = \Phi^{-1}\left( 1 - \frac{\alpha} 2 \right),
\label{CriticalSimple2}
\end{align}
when computing the confidence interval~(\ref{SimpleConfidence}).

\subsubsection*{Critical Value for the Case $a_1 \ll a_2 \ll u$}

Analogous to the discussion of (\ref{CriticalSimple}), the critical value (%
\ref{CriticalSimple2}) is again potentially conservative because it is based
on the case where two of the inequalities $m(\theta_0) \leq 0$ (for either
the lower or the upper bound, respectively) are jointly binding.\footnote{%
It is also conservative, because the information in the correlation matrix $%
\Sigma_m$ is not used to construct~(\ref{CriticalSimple2}). It corresponds
to the the most extreme case where both lower bound estimators $\widehat a_1$
and $\widehat a_2$ (or both upper bound estimators $\widehat a_3$ and $%
\widehat a_4$) are perfectly negatively correlated.} For example, if we find
that $a_1 \ll a_2 \ll u$ (by which we again mean that the null hypotheses $%
H_0:a_1 = a_2$, vs. $H_a:a_1<a_2$, and $H_0:a_2 = u$, vs. $H_a:a_2<u$, are
rejected with very high confidence), then a natural confidence interval to
report is
\begin{align*}
\widehat \Theta &= \left[ \widehat a_2 - \frac{\Phi^{-1}\left( 1 - \alpha
\right) \widehat \Sigma_{a,22}^{1/2} } {\sqrt{n}} , \min \left( \widehat a_3
+ \frac{\Phi^{-1}\left( 1 - \frac{\alpha} 2 \right) \widehat
\Sigma_{a,33}^{1/2} } {\sqrt{n}} , \widehat a_4 + \frac{\Phi^{-1}\left( 1 -
\frac{\alpha} 2 \right) \widehat \Sigma_{a,44}^{1/2} } {\sqrt{n}} \right) %
\right] .
\end{align*}
Note that the lower bound of $\widehat \Theta$ now corresponds to inverting
a standard one-sided t-test. Analogous confidence intervals can obviously be
constructed in other cases, e.g. $\ell \ll a_3 \ll a_4$ or $a_2 \ll a_1 \ll
a_4 \ll a_3$, etc.

The different critical values and corresponding confidence intervals
discussed above correspond to cases where different subsets of the
inequalities $m(\theta_0) \leq 0$ can be simultaneously binding, i.e.~to a
moment selection problem. A much more general discussion of moment selection
is given e.g.~in \citeasnoun{AndrewsSoares2010}. Different confidence
intervals than those discussed here, e.g.~based on different objective
functions $\widehat Q(\theta)$, can of course also be considered.

It should be noted that pretesting is not required if we use the approach in %
\citeasnoun{Hahn2014a} who obtain a confidence interval by inverting the
Likelihood Ratio test for the composite null and composite alternative test.
Their current results do not cover the case considered here and we did not
attempt the non-trivial extension to the case considered here.

\section{Application to the Illinois bonus experiment}

\label{s:application}

\subsection{The re-employment bonus experiment}

In 1984, the Illinois Department of Employment
Security conducted a randomized social experiment.\footnote{%
The population consisted of those who filed for UI between July 29, 1984 and November 17, 1984. A complete description of the experiment and a summary of its results can be
found in \citeasnoun{Woodbury1987}.} The goal of the experiment was to
explore, whether re-employment bonuses paid to Unemployment Insurance (UI)
beneficiaries (treatment 1) or their employers (treatment 2) reduced the
length of unemployment spells.

Both treatments consisted of a \$500 re-employment bonus, which was about
four times the average weekly unemployment insurance benefit. In the
experiment, newly unemployed UI claimants were randomly divided into three
groups: \newline
1. The \emph{Claimant Bonus Group}. The members of this group were
instructed that they would qualify for a cash bonus of \$500 if they found a
job (of at least 30 hours) within 11 weeks and, if they held that job for at
least 4 months. A total of 4186 individuals were selected for this group,
and 3527 (84\%) agreed to participate. \newline
2. The \emph{Employer Bonus Group}. The members of this group were told that
their next employer would qualify for a cash bonus of \$500 if they, the
claimants, found a job (of at least 30 hours) within 11 weeks and, if they
held that job for at least four months. A total of 3963 were selected for
this group and 2586 (65\%) agreed to participate. \newline
3. The \emph{Control Group}, i.e. all claimants not assigned to one of the
treatment groups. This group consisted of 3952 individuals. The individuals
assigned to the control group were excluded from participation in the
experiment. In fact, they did not know that the experiment took place.

The descriptive statistics in Table 2 in \citeasnoun{Woodbury1987} confirm
that the randomization resulted in three similar groups.

\subsection{Results of previous studies}

\citeasnoun{Woodbury1987} concluded from a direct comparison of the control
group and the two treatment groups that the claimant bonus group had a
significantly shorter average unemployment duration. The average
unemployment duration was also shorter for the employer bonus group, but the
difference was not significantly different from zero. In Illinois UI
benefits end after 26 weeks and since administrative data were used, all
unemployment durations are censored at 26 weeks. Woodbury and Spiegelman
ignore the censoring and take as outcome variable the number of weeks of
insured unemployment.

\citeasnoun{Meyer1996} analyzed the same data but focused on the treatment
effects on conditional transition probabilities which allows him to properly
account for censoring. Meyer focuses on the conditional transitions rates
because both labor supply and search theory imply specific dynamic treatment
effects. The bonus is only given to an unemployed individual if (s)he finds
a job within 11 weeks and retains it for four months. The cash bonus is the
same for all unemployed. Theory predicts that (i) the transition rate during
the eligibility period (first 11 weeks) will be higher in the two treatment
groups compared with the control group, and (ii) that the transition rate in
the treatment groups will rise just before the end of the eligibility
period, as the unemployed run out of time to collect the bonus.

To test these predictions, \citeasnoun{Meyer1996} estimates a proportional
hazard (PH) model with a flexible specification of the baseline hazard. He
uses the treatment indicator as an explanatory variable. Since there was
partial compliance with treatment his estimator can be interpreted as a
intention to treat (ITT) estimator.\footnote{%
The partial compliance is addressed in detail by \citeasnoun{Bijwaard2005}.
They introduce a new method to handle the selective compliance in the
treatment group. If there is full compliance in the control group, their
two-stage linear rank estimator is able to handle the selective compliance
in the treatment group even for censored durations. In order to achieve this
they assume a MPH structure for the transition rate. Their estimates
indicate that the ITT estimates by \citeasnoun{Meyer1996} underestimate the
true treatment effect.} In his analysis Meyer controls for age, the
logarithm of base period earnings, ethnicity , gender and the logarithm of
the size of the UI benefits. He finds a significantly positive effect of the
claimant bonus and a positive but insignificant effect of the employer
bonus. A more detailed analysis of the effects for the claimant group
reveals a positive effect on the transition rate during the first 11 weeks
in unemployment, an increased effect during week 9 and 10, and no
significant effect on the transition rate after week 11 as predicted by
labor supply and search theory.

\subsection{Estimates of bounds}

In his study \citeasnoun{Meyer1996} relies on the proportionality of the
hazard rate to investigate his hypotheses. We now ask what can be said if
the assumptions of the MPH model do not hold, that is what can be identified
if we rely solely on random assignment and the additional assumptions.
As in \citeasnoun{Meyer1996} we consider the ITT effect, that is, we do not correct for partial
compliance. We divide the 24 month observation period into 12 subperiods:
week 1-2, week 3-4, ... , week 23-24. The reason for this is that there is a
pronounced even-odd week effect in the data, with higher transition rate
during odd weeks. With these subperiods the predictions we wish to test are:
(i) a positive treatment effect during periods 1-5, i.e.
\begin{equation*}
\mathrm{ATETS}_{t}>0\ ,\; \;t=1,\ldots ,5,
\end{equation*}%
(ii) no effect after the bonus offer has expired in periods 6-12, i.e.
\begin{equation*}
\mathrm{ATETS}_{t}=0\ ,\; \;t=6,\ldots ,12,
\end{equation*}%
and (iii) a larger effect of the bonus offer at the end of the eligibility
period in period 5, i.e.
\begin{equation*}
\mathrm{ATETS}_{5}>\mathrm{ATETS}_{4}.
\end{equation*}%
Note that in this experiment the treatment assignment is in period 1, so
that in $\mathrm{ATETS}_{t}$ the superscripts 1 and 0 are $t$ vectors with
components equal to 1 and 0.

We report both the bounds that are obtained by simply replacing the
population moments with their sample analogs, as well as the confidence
intervals based on the approach described in section \ref{s:inference}.%
\footnote{%
The covariance matrix $\Sigma _{a}$ is estimated using the bootstrap with
399 replications. Constructing confidence intervals furthermore requires
moment selection, e.g.~for the bounds under just random assignment we find
that with very high confidence only one inequality is binding for the lower
as well as the upper bound. Details are available from the authors upon
request.} Table \ref{t:bounds1} presents the upper and the lower bound and
the confidence interval on \textrm{ATETS}$_{t}$ for the claimant group
assuming only random assignment. We find that the instantaneous treatment
effect on the transition probability (week 1-2) is point identified and
indicates a positive effect of the re-employment bonus. The transition
probability is about 2 percentage points higher in the treatment group
compared to the control group. This estimate is statistically significant.
From week 3-4 and onwards the bounds are quite wide. In fact, without
further assumptions we cannot rule out that the bonus actually has a
negative impact on the conditional transition probability after week 3.
However, the bounds are nevertheless informative on the average treatment
effect in all time periods.

Table \ref{t:bounds1} also shows that the confidence intervals are
marginally wider than the actual bounds. That is the uncertainty arising
from the dynamic selection is far greater than the uncertainty due to
sampling variation.

Next, Table \ref{t:bounds1} presents bounds under the additional assumptions
in Section 4. As expected, if we impose additional assumptions the bounds
are considerably narrower. Under MTR and CS we can rule out very large
negative and very large positive dynamic treatment effects. Imposing MTR, CS
as well as PCO further tightens the bounds. If these assumptions hold
simultaneously we can, if we disregard sampling variation, rule out that the
bonus offer has a negative effect on the transition rate out of unemployment
up to week 20. This conclusions changes slightly when sampling variation is
taken into account.

Let us return to the three hypotheses suggested by labor supply and search
theory, and consider our most restrictive bounds under MTR, CS and PCO. We
find that there is a positive effect of the bonus offer on the conditional
transition rate up to week 11. This confirms the first hypothesis. The upper
bound increases in time period 5 (weeks 9-10), but the lower bound does not
increase enough, so that both an increase and no change (and even a small
decrease) in the transition probability out of unemployment are consistent
with the data. Now consider the third hypothesis that there is no effect on
the transition rate after week 11. Again the bounds do not rule out that
there is a positive effect on the conditional transition probability after
week 11. These results illustrate that the evidence for the second and third
hypotheses presented by a number of authors rely on the imposed structure,
e.g. proportionality of the hazard or the restrictions implied by a
particular discrete-time duration model.

We next examine heterogenous effects. To this end we split our sample by
gender, race and pre-unemployment income and estimate our bounds for each
subgroup. We provide results for bounds without additional assumptions and
bounds under MTR, CS and PCO. The other bounds are available upon request.
If we focus on the bounds under MTR, CS and PCO, Table \ref{t:bounds_gender}
indicates several interesting differences between males and females. For
males we find significant effects in the beginning of the unemployment spell
(weeks 1-2) and shortly before the bonus expires (weeks 7-10). For females
on the other hand we only find significant effects in weeks 1-4, but no
effects in weeks 5-11. This indicates that females quickly responds to the
bonus offer, whereas a large part of the effects for males occur shortly
before the end of the subsidy. Table \ref{t:bounds_race} in Appendix B also
reveals some differences between blacks and non-blacks. For both groups we
find significant effects during the first 11 weeks of unemployment, but for
non-blacks the bonus offer also increases the transition rates after the
bonus offers has expired (e.g. during weeks 15-16). Finally, Tables \ref%
{t:bounds_income} in Appendix B reveals no significant differences between
how workers with low and high income react to the bonus offer.

\section{Conclusions}

\label{s:conclusions}

In this paper, we have derived bounds on treatment effects on conditional
transition probabilities under (sequential) randomization. The partial
identification problem arises since random assignment only ensures
comparability of the treatment and control groups at the time of
randomization. In the literature this problem is often refereed to as the
dynamic selection problem. For that reason only instantaneous or short-run
effects are point identified, whereas dynamic or long-run effects in general
are not point identified. Our weakest bounds impose no assumptions beyond
(sequential) random assignment, so that they are not sensitive to arbitrary
functional form assumptions, require no additional covariates and allow
arbitrary heterogenous treatment effects as well as arbitrary unobserved
heterogeneity. These non-parametric bounds offer an alternative to
semi-parametric methods. They tend to be wide and therefore we have also
derived more informative bounds under additional assumptions that often hold
in semi-parametric reduced form and structural models.

An analysis of data from the Illinois re-employment bonus experiment shows
that our bounds are informative about average treatment effects. It also
demonstrates that previous results on the evolution of the average treatment
effect require assumptions such as the proportionality of the hazard rate or
those embodied in a particular (semi-)parametric discrete-time hazard model.


\newpage

\ifx\undefined\BySame
\newcommand{\BySame}{\leavevmode\rule[.5ex]{3em}{.5pt}\ }
\fi
\ifx\undefined\textsc
\newcommand{\textsc}[1]{{\sc #1}}
\newcommand{\emph}[1]{{\em #1\/}}
\let\tmpsmall\small
\renewcommand{\small}{\tmpsmall\sc}
\fi

\newpage

\section*{Tables}

\begin{table}[h!]
\caption{{\protect \small For the case $T=2$ there are sixteen possible
realizations for the potential outcomes $Y_{1}^{1}$, $Y_{2}^{1}$, $Y_{1}^{0}$%
, $Y_{2}^{0}$, and the corresponding probabilities are given in the table.
The table also shows the row- and column-sums that are point identified from
the data (``observable''). The four underlined probabilities are those that enter into
the numerator of $\mathbb{E} \left(Y_{2}^0 \big| Y_{1}^{1}=0 \right)$, see
equation \eqref{InequT2e1} in the main text.}}
\label{tab:ProbT2}
\begin{center}
\begin{tabular}{c|@{\,}c@{\;}c@{}c@{}c|@{\,}l}
& {\footnotesize $Y^0 = (0,0)$} & {\footnotesize $Y^0 = (0,1)$} &
{\footnotesize $Y^0 = (1,0)$} & {\footnotesize $Y^0 = (1,1)$} & observable
(row sum) \\ \hline
&  &  &  &  &  \\[-10pt]
{\footnotesize $Y^1 = (0,0)$} & $p_{00,00}$ & $\underline{p_{00,01}}$ & $p_{00,10}$
& $\underline{p_{00,11}}$ & $\big \} \,$ {\footnotesize $\Pr \left( Y=(0,0) \big| %
D=1\right)$} \\[5pt]
{\footnotesize $Y^1 = (0,1)$} & $p_{01,00}$ & $\underline{p_{01,01}}$ & $p_{01,10}$
& $\underline{p_{01,11}}$ & $\big \} \,$ {\footnotesize $\Pr \left( Y=(0,1) \big| %
D=1\right)$} \\[5pt]
{\footnotesize $Y^1 = (1,0)$} & $p_{10,00}$ & $p_{10,01}$ & $p_{10,10}$ & $%
p_{10,11}$ & \multirow{2}{*}{$\bigg \} \,$ \footnotesize $ \Pr \left( Y_1 =
1 \big| D=1\right)$} \\[5pt]
{\footnotesize $Y^1 = (1,1)$} & $p_{11,00}$ & $p_{11,01}$ & $p_{11,10}$ & $%
p_{11,11}$ &  \\ \hline
&  &  &  &  &  \\[-20pt]
& $\underbrace{\phantom{?????}}$ & $\underbrace{\phantom{?????}}$ &
\multicolumn{2}{c}{$\underbrace{\phantom{?????????????????}}$} &  \\
observable & {\footnotesize $\Pr \left( Y=(0,0) \big| D=0\right)$} &
{\footnotesize $\Pr \left( Y=(0,1) \big| D=0\right)$} & \multicolumn{2}{c}%
{\footnotesize $\Pr \left( Y_1 = 1 \big| D=0\right)$} &    \\[-15pt]
&  &  &  &  &  \\
(col.~sum) &  &  &  &
\end{tabular}%
\end{center}
\end{table}

\clearpage \newpage

\begin{table}[h!]
\caption{Bounds on $ATETS^{1,0}$ for the Illinois job bonus experiment}
\label{t:bounds1}\centering{\small
\begin{tabularx}{\textwidth}{lzzzzzzzz}
\toprule
&     \multicolumn{4}{c}{No assumption bounds [A]} & \multicolumn{4}{c}{MTR+CS [B]}   \\
\cmidrule(r){2-5}
\cmidrule(l){6-9}
& Lower-CI &  LB & UB & Upper-CI & Lower-CI &  LB & UB & Upper-CI \\
& (1) & (2) & (3) & (4) & (5) & (6) & (7) & (8) \\
\hline
   Week &        \\
       1-2    & ~0.012 & ~0.023 & 0.023& 0.034    &~0.012  & ~0.023  &0.023 &0.034               \\
       3-4    &-0.145& -0.137& 0.094& 0.102       &~0.000  & ~0.011  &0.038 &0.050        \\
       5-6    &-0.259& -0.251& 0.074& 0.082       &-0.007 & ~0.004  &0.046 &0.056          \\
       7-8    &-0.346& -0.339& 0.078& 0.086       &~0.004  & ~0.013  &0.063 &0.073        \\
       9-10   &-0.452& -0.444& 0.069& 0.077       &~0.000  & ~0.008  &0.069 &0.079         \\
       11-12  &-0.552& -0.544& 0.062& 0.070       &~0.000  & ~0.008  &0.062 &0.072        \\
       13-14  &-0.655& -0.648& 0.056& 0.064       &-0.010 & -0.002 &0.056 &0.064         \\
       15-16  &-0.750& -0.743& 0.051& 0.058       &-0.004 & ~0.003  &0.051 &0.058         \\
       17-18  &-0.844& -0.836& 0.049& 0.057       &-0.007 & ~0.000  &0.049 &0.057         \\
       19-20  &-0.943& -0.936& 0.049& 0.057       &-0.011 & -0.004 &0.049 &0.056         \\
       21-22  &-0.994& -0.953& 0.047& 0.056       &-0.028 & -0.021 &0.047 &0.055         \\
       23-24  &-0.989& -0.944& 0.056& 0.064       &-0.011 & -0.002 &0.056 &0.064         \\
\hline
&     \multicolumn{4}{c}{PCO [C]} & \multicolumn{4}{c}{MTR+CS+PCO [D]}   \\
\cmidrule(r){2-5}
\cmidrule(l){6-9}
& Lower-CI &  LB & UB & Upper-CI & Lower-CI &  LB & UB & Upper-CI \\
& (1) & (2) & (3) & (4) & (5) & (6) & (7) & (8) \\
\hline
   Week &        \\
       1-2    & ~0.012& ~0.023 &0.023 &0.034         &~0.012  &~0.023  &0.023 &0.034          \\
       3-4    &-0.131& -0.123& 0.094& 0.102          &~0.002  &~0.014  &0.038 &0.049   \\
       5-6    &-0.209& -0.202& 0.074& 0.082          &-0.004 &~0.007  &0.046 &0.055     \\
       7-8    &-0.256& -0.247& 0.078& 0.087          &~0.008  &~0.016  &0.063 &0.072   \\
       9-10   &-0.306& -0.299& 0.069& 0.077          &~0.004  &~0.012  &0.069 &0.078    \\
       11-12  &-0.348& -0.340& 0.062& 0.070          &~0.004  &~0.012  &0.062 &0.071   \\
       13-14  &-0.388& -0.379& 0.056& 0.064          &-0.004 &~0.003  &0.056 &0.064   \\
       15-16  &-0.419& -0.411& 0.051& 0.058          & 0.000 &~0.007  &0.051 &0.059    \\
       17-18  &-0.445& -0.438& 0.049& 0.057          &-0.003 &~0.005  &0.049 &0.058    \\
       19-20  &-0.472& -0.464& 0.049& 0.057          &-0.006 &~0.001  &0.049 &0.057   \\
       21-22  &-0.504& -0.496& 0.047& 0.063          &-0.022 &-0.014 &0.047 &0.055    \\
       23-24  &-0.523& -0.513& 0.056& 0.073          &-0.006 &~0.003  &0.056 &0.065   \\
\bottomrule
\multicolumn{9}{p{0.98\textwidth}}{\footnotesize Notes: CI is 95\% confidence intervals. Variances and covariances used to obtain the CI are estimated using bootstrap (399 replications).}
\end{tabularx}
}
\end{table}

\clearpage \newpage

\begin{table}[h!]
\caption{Bounds on $ATETS^{1,0}$ for the Illinois job bonus experiment.
Heterogenous effects for males and females}
\label{t:bounds_gender}\centering{\small
\begin{tabularx}{\textwidth}{lzzzzzzzz}
\toprule
&     \multicolumn{8}{c}{\textbf{Panel A: Males}}    \\
&     \multicolumn{4}{c}{No assumption bounds} & \multicolumn{4}{c}{MTR+CS+PCO}   \\
\cmidrule(r){2-5}
\cmidrule(l){6-9}
& Lower-CI &  LB & UB & Upper-CI & Lower-CI &  LB & UB & Upper-CI \\
& (1) & (2) & (3) & (4) & (1) & (2) & (3) & (4) \\
\hline
   Week &        \\
       1-2  & -0.004& ~0.016 &0.016 &0.037    &~0.002 &~0.016 &0.016 &0.030            \\
       3-4  & -0.152& -0.141& 0.094& 0.105    &-0.004 &~0.009 &0.026 &0.039    \\
       5-6  & -0.269& -0.259& 0.075& 0.084    &-0.010 &~0.003 &0.030 &0.043     \\
       7-8  & -0.349& -0.338& 0.085& 0.096    &~0.009 &~0.024 &0.054 &0.069     \\
       9-10 & -0.464& -0.453& 0.076& 0.087    &~0.000 &~0.014 &0.070 &0.084     \\
       11-12& -0.573& -0.562& 0.069& 0.080    &~0.005 &~0.015 &0.069 &0.081     \\
       13-14& -0.688& -0.676& 0.065& 0.076    &-0.004 &~0.006 &0.065 &0.077   \\
       15-16& -0.793& -0.782& 0.054& 0.064    &~0.004 &~0.014 &0.054 &0.064    \\
       17-18& -0.899& -0.887& 0.056& 0.067    &-0.008 &~0.003 &0.056 &0.066   \\
       19-20& -0.994& -0.941& 0.059& 0.071    &-0.004 &~0.008 &0.059 &0.071   \\
       21-22& -1.006& -0.948& 0.052& 0.063    &-0.028 &-0.017 &0.052 &0.066 \\
       23-24& -1.006& -0.941& 0.059& 0.071    &-0.010 &~0.002 &0.059 &0.074   \\

\hline
&     \multicolumn{4}{c}{PCO [C]} & \multicolumn{4}{c}{MTR+CS+PCO [D]}   \\
\cmidrule(r){2-5}
\cmidrule(l){6-9}
&     \multicolumn{8}{c}{\textbf{Panel B: Females}}    \\
&     \multicolumn{4}{c}{No assumption bounds} & \multicolumn{4}{c}{MTR+CS+PCO}   \\
& (1) & (2) & (3) & (4) & (1) & (2) & (3) & (4) \\
\hline
   Week &        \\
       1-2    &~0.008 &~0.031 &0.031 &0.054    &~0.014 &~0.031 &0.031 &0.047               \\
       3-4    &-0.143 &-0.131 &0.093 &0.105    &~0.003 &~0.019 &0.053 &0.069        \\
       5-6    &-0.251 &-0.239 &0.074 &0.085    &~0.000 &~0.012 &0.066 &0.080        \\
       7-8    &-0.348 &-0.337 &0.068 &0.079    &-0.006 &~0.005 &0.068 &0.082       \\
       9-10   &-0.441 &-0.430 &0.060 &0.071    &-0.003 &~0.009 &0.060 &0.073        \\
       11-12  &-0.528 &-0.517 &0.053 &0.064    &-0.002 &~0.008 &0.053 &0.066       \\
       13-14  &-0.616 &-0.606 &0.045 &0.055    &-0.011 &~0.000 &0.045 &0.055     \\
       15-16  &-0.698 &-0.686 &0.046 &0.057    &-0.012 &~0.000 &0.046 &0.059      \\
       17-18  &-0.775 &-0.764 &0.041 &0.052    &-0.008 &~0.007 &0.041 &0.055       \\
       19-20  &-0.861 &-0.851 &0.036 &0.047    &-0.016 &-0.006 &0.036 &0.047     \\
       21-22  &-0.949 &-0.936 &0.041 &0.054    &-0.022 &-0.011 &0.041 &0.055     \\
       23-24  &-1.020 &-0.948 &0.052 &0.066    &-0.009 &~0.004 &0.052 &0.068      \\
\bottomrule
\multicolumn{9}{p{0.98\textwidth}}{\footnotesize Notes: CI is 95\% confidence intervals. Variances and covariances used to obtain the CI are estimated using bootstrap (399 replications).}
\end{tabularx}
}
\end{table}

\newpage \addcontentsline{toc}{section}{Appendix A}

\section*{Appendix A: Proofs}

\renewcommand{\theequation}{A.\arabic{equation}} \setcounter{equation}{0}

\noindent \textbf{Proof of Theorem \ref{ATETSBasBounds-t}}\newline

\noindent We use the following notation for the distribution of the
potential outcomes. For $d=0,1$
\begin{align*}
p_{t}^{d}(1|0,0)& =:\Pr (Y_{t}^{d}=1|\overline{Y}_{t-1}^{1}=0,\overline{Y}%
_{t-1}^{0}=0), \\
p_{t}^{d}(1|0,\neq 0)& =:\Pr (Y_{t}^{d}=1|\overline{Y}_{t-1}^{1}=0,\overline{%
Y}_{t-1}^{0}\neq 0), \\
p_{t}^{d}(1|\neq 0,0)& =:\Pr (Y_{t}^{d}=1|\overline{Y}_{t-1}^{1}\neq 0,%
\overline{Y}_{t-1}^{0}=0),
\end{align*}%
and for the joint distribution of $\overline{Y}_{t-1}^{1},\overline{Y}%
_{t-1}^{0}$
\begin{align*}
p_{t-1}(0,0)& =:\Pr (\overline{Y}_{t-1}^{1}=0,\overline{Y}_{t-1}^{0}=0), \\
p_{t-1}(0,\neq 0)& =:\Pr (\overline{Y}_{t-1}^{1}=0,\overline{Y}%
_{t-1}^{0}\neq 0), \\
p_{t-1}(\neq 0,0)& =:\Pr (\overline{Y}_{t-1}^{1}\neq 0,\overline{Y}%
_{t-1}^{0}=0),
\end{align*}%
We derive bounds on $\mathrm{ATETS}_{t}$ defined by
\begin{equation}
\mathbb{E}\left[ Y_{t}^{1}|\overline{Y}_{t-1}^{1}=0\right] -\mathbb{E}\left[
Y_{t}^{0}|\overline{Y}_{t-1}^{1}=0\right]  \label{eqATETS}
\end{equation}%
with the data providing the observed transition probabilities $\Pr
(Y_{t}=y_{t}|\overline{Y}_{t-1}=0,D=1)$ and $\Pr (Y_{t}=y_{t}|\overline{Y}%
_{t-1}=0,D=0)$.

Under Assumption 1
\begin{equation*}
\mathbb{E}[Y_{t}^{1}|\overline{Y}_{t-1}^{1}=0]=\Pr (Y_{t}=1|\overline{Y}%
_{t-1}=0,D=1),
\end{equation*}%
so that if $\Pr (\overline{Y}_{t-1}^{1}=0|D=1)=\Pr (\overline{Y}%
_{t-1}=0|D=1)>0$ then $\mathbb{E}[Y_{t}^{1}|\overline{Y}_{t-1}^{1}=0]$ is
point-identified, and if $\Pr (\overline{Y}_{t-1}^{1}=0|D=1)=\Pr (\overline{Y%
}_{t-1}=0|D=1)=0$ then $\mathbb{E}[Y_{t}^{1}|\overline{Y}_{t-1}^{1}=0],%
\mathbb{E}[Y_{t}^{0}|\overline{Y}_{t-1}^{1}=0]$ and $\mathrm{ATETS}_{t}$ are
not defined. Note that the point identification of this mean is similar to
the point identification of the treated mean in the \textrm{ATET} in static
settings.

Next, we have for the counterfactual transition probability
\begin{equation}
\mathbb{E}\left[ Y_{t}^{0}|\overline{Y}_{t-1}^{1}=0\right] =\frac{%
p_{t}^{0}(1|0,0)p_{t-1}(0,0)+p_{t}^{0}(1|0,\neq 0)p_{t-1}(0,\neq 0)}{%
p_{t-1}(0,0)+p_{t-1}(0,\neq 0)}.  \label{teqbas00}
\end{equation}%
By Assumption 1
\begin{equation*}
\Pr (Y_{t}=1,\overline{Y}_{t-1}=0|D=0)=\Pr (Y_{t}^{0}=1,\overline{Y}%
_{t-1}^{0}=0|D=0)=\Pr (Y_{t}^{0}=1,\overline{Y}_{t-1}^{0}=0).
\end{equation*}%
By the law of total probability
\begin{equation*}
\Pr (Y_{t}^{0}=1,\overline{Y}_{t-1}^{0}=0)=\Pr (\overline{Y}%
_{t-1}^{1}=0,Y_{t}^{0}=1,\overline{Y}_{t-1}^{0}=0)+\Pr (\overline{Y}%
_{t-1}^{1}\neq 0,Y_{t}^{0}=1,\overline{Y}_{t-1}^{0}=0)=
\end{equation*}%
\begin{equation*}
p_{t}^{0}(1|0,0)p_{t-1}(0,0)+p_{t}^{0}(1|\neq 0,0)p_{t-1}(\neq 0,0).
\end{equation*}%
Therefore,
\begin{equation*}
\Pr (Y_{t}=1,\overline{Y}%
_{t-1}=0|D=0)=p_{t}^{0}(1|0,0)p_{t-1}(0,0)+p_{t}^{0}(1|\neq 0,0)p_{t-1}(\neq
0,0)
\end{equation*}%
Solving for $p_{t}^{0}(1|0,0)$ gives
\begin{equation*}
p_{t}^{0}(1|0,0)=\frac{\Pr (Y_{t}=1,\overline{Y}_{t-1}=0|D=0)-p_{t}^{0}(1|%
\neq 0,0)p_{t-1}(\neq 0,0)}{p_{t-1}(0,0)}.
\end{equation*}%
and upon substitution
\begin{equation*}
\mathbb{E}\left[ Y_{t}^{0}|\overline{Y}_{t-1}^{1}=0\right] =\frac{\Pr
(Y_{t}=1,\overline{Y}_{t-1}=0|D=0)}{p_{t-1}(0,0)+p_{t-1}(0,\neq 0)}-\frac{%
p_{t}^{0}(1|\neq 0,0)p_{t-1}(\neq 0,0)-p_{t}^{0}(1|0,\neq 0)p_{t-1}(0,\neq 0)%
}{p_{t-1}(0,0)+p_{t-1}(0,\neq 0)}.
\end{equation*}

The expression on the right-hand side is decreasing in $p_{t}^{0}(1|\ne0 ,0)$
and increasing in $p_{t}^{0}(1|0,\ne0)$. The lower bound is obtained by
setting $p_{t}^{0}(1|\ne0 ,0)$ at 1 and $p_{t}^{0}(1|0,\ne0)$ at 0 and the
upper bound by setting $p_{t}^{0}(1|\ne0,0)$ at 0 and $p_{t}^{0}(1|0, \ne0)$
at 1 so that
\begin{equation*}
\frac{\Pr(Y_{t}=1|\overline{Y}_{t-1}=0, D=0)\Pr(\overline{Y}_{t-1}=0| D=0)
-p_{t-1}(\ne0,0)}{p_{t-1}(0,0)+ p_{t-1}(0,\ne0)}
\end{equation*}
\begin{equation*}
\leq \mathbb{E}\left[ Y_{t}^{0}|\overline{Y}_{t-1}^{1}=0\right] \leq
\end{equation*}%
\begin{equation*}
\frac{\Pr(Y_{t}=1|\overline{Y}_{t-1}=0, D=0)\Pr(\overline{Y}_{t-1}=0| D=0)
+p_{t-1}(0,\ne0)}{p_{t-1}(0,0)+ p_{t-1}(0,\ne0)}.
\end{equation*}
where we note that
\begin{equation*}
\Pr(Y_{t}=1|\overline{Y}_{t-1}=0, D=0)\Pr(\overline{Y}_{t-1}=0| D=0)=\Pr
(Y_{t}=1, \overline{Y}_{t-1}=0| D=0)=0
\end{equation*}
if $\Pr(\overline{Y}_{t-1}=0| D=0)=0$.

Because
\begin{equation*}
\Pr (\overline{Y}_{t-1}=0|D=1)=p_{t-1}(0,0)+p_{t-1}(0,\neq 0)
\end{equation*}%
and
\begin{equation*}
\Pr (\overline{Y}_{t-1}=0|D=0)=p_{t-1}(0,0)+p_{t-1}(\neq 0,0)
\end{equation*}%
we have
\begin{equation}
\frac{\lbrack \Pr (Y_{t}=1|\overline{Y}_{t-1}=0,D=0)-1]\Pr \left( \overline{Y%
}_{t-1}=0|D=0\right) +p_{t-1}(0,0)}{\Pr (\overline{Y}_{t-1}=0|D=1)}
\label{ATETSt2}
\end{equation}%
\begin{equation*}
\leq \mathbb{E}\left[ Y_{t}^{0}|\overline{Y}_{t-1}^{1}=0\right] \leq
\end{equation*}%
\begin{equation*}
\frac{\Pr (Y_{t}=1|\overline{Y}_{t-1}=0,D=0)\Pr \left( \overline{Y}%
_{t-1}=0|D=0\right) -p_{t-1}(0,0)}{\Pr (\overline{Y}_{t-1}=0|D=1)}+1.
\end{equation*}%
The upper bound is decreasing and the lower bound is increasing in $%
p_{t-1}(0,0)$. By the Bonferroni inequality
\begin{equation*}
p_{t-1}(0,0)\geq \max \left \{ \Pr (\overline{Y}_{t-1}^{1}=0)+\Pr (\overline{%
Y}_{t-1}^{0}=0)-1,0\right \} =
\end{equation*}%
\begin{equation*}
\max \left \{ \Pr \left( \overline{Y}_{t-1}=0|D=1\right) +\Pr \left(
\overline{Y}_{t-1}=0|D=0\right) -1,0\right \} .
\end{equation*}

If
\begin{equation*}
\Pr \left( \overline{Y}_{t-1}=0|D=1\right) +\Pr \left( \overline{Y}%
_{t-1}=0|D=0\right) -1\leq 0
\end{equation*}%
the lower bound on $p_{t-1}(0,0)$ is 0. In that case the lower bound in (\ref%
{ATETSt2}) is non-positive and the upper bound is greater than or equal to 1
so that
\begin{equation*}
0\leq \mathbb{E}\left[ Y_{t}^{0}|\overline{Y}_{t-1}^{1}=0\right] \leq 1.
\end{equation*}%
If $\Pr \left( \overline{Y}_{t-1}=0|D=1\right) +\Pr \left( \overline{Y}%
_{t-1}=0|D=0\right) -1>0$ we have upon substitution of the lower bound on $%
p_{t-1}(0,0)$ into (\ref{ATETSt2}) and because the probability $\mathbb{E}%
\left[ Y_{t}^{0}|\overline{Y}_{t-1}^{1}=0\right] $ is bounded by zero and
one
\begin{equation*}
\max \left \{ 0,\frac{\Pr (Y_{t}=1|\overline{Y}_{t-1}=0,D=0)\Pr \left(
\overline{Y}_{t-1}=0|D=0\right) -1}{\Pr (\overline{Y}_{t-1}=0|D=1)}+1\right
\}
\end{equation*}%
\begin{equation}
\leq \mathbb{E}\left[ Y_{t}^{0}|\overline{Y}_{t-1}^{1}=0\right] \leq
\label{teqsolmaxmin11}
\end{equation}%
\begin{equation*}
\min \left \{ 1,\frac{1-[1-\Pr (Y_{t}=1|\overline{Y}_{t-1}=0,D=0)]\Pr \left(
\overline{Y}_{t-1}=0|D=0\right) }{\Pr (\overline{Y}_{t-1}=0|D=1)}\right \} .
\end{equation*}%
Finally, we combine these bounds with the point-identified $\mathbb{E}%
[Y_{t}^{1}|\overline{Y}_{t-1}^{1}=0]$ to obtain bounds on $\mathrm{ATETS}%
_{t} $.\newline

\noindent \textbf{Proof of Theorem \ref{ATETSBoundsMTR-t}} \newline

\noindent As above, under Assumption 1 $\mathbb{E}[Y_{t}^{1}|\overline{Y}%
_{t-1}^{1}=0]=\Pr (Y_{t}=1|\overline{Y}_{t-1}=0,D=1),$ so that if $\Pr (%
\overline{Y}_{t-1}=0|D=1)>0$ then $\mathbb{E}[Y_{t}^{1}|\overline{Y}%
_{t-1}^{1}=0]$ is point-identified, and if $\Pr (\overline{Y}_{t-1}=0|D=1)=0$
then $\mathrm{ATETS}_{t}$ is not defined. If $\Pr (\overline{Y}%
_{t-1}=0|D=1)>0$ we have from (\ref{ATETSt2})

\begin{equation}
\frac{\lbrack \Pr (Y_{t}=1|\overline{Y}_{t-1}=0,D=0)-1]\Pr \left( \overline{Y%
}_{t-1}=0|D=0\right) +p_{t-1}(0,0)}{\Pr (\overline{Y}_{t-1}=0|D=1)}
\label{ATETSt2_MTR}
\end{equation}%
\begin{equation*}
\leq \mathbb{E}\left[ Y_{t}^{0}|\overline{Y}_{t-1}^{1}=0\right] \leq
\end{equation*}%
\begin{equation*}
\frac{\Pr (Y_{t}=1|\overline{Y}_{t-1}=0,D=0)\Pr \left( \overline{Y}%
_{t-1}=0|D=0\right) -p_{t-1}(0,0)}{\Pr (\overline{Y}_{t-1}=0|D=1)}+1.
\end{equation*}%
Because the lower bound is increasing in $p_{t-1}(0,0)$ and the upper bound
decreasing in $p_{t-1}(0,0)$ we need the lower bound on this probability. We
have
\begin{equation*}
p_{t-1}(0,0)=\Pr (Y_{t-1}^{1}=0,\ldots ,Y_{1}^{1}=0,Y_{t-1}^{0}=0,\ldots
,Y_{1}^{0}=0)=
\end{equation*}%
\begin{equation*}
\Pr (Y_{t-1}^{1}=0,Y_{t-1}^{0}=0|S_{t-2})\Pr (Y_{t-2}^{1}=0,\ldots
,Y_{1}^{1}=0,Y_{t-2}^{0}=0,\ldots ,Y_{1}^{0}=0).
\end{equation*}%
By Assumption \ref{MTR-gen} either
\begin{equation}
\Pr \left( Y_{t-1}^{1}=0|S_{t-2},V\right) \leq \Pr \left(
Y_{t-1}^{0}=0|S_{t-2},V\right) ,  \label{eq1a}
\end{equation}%
or
\begin{equation}
\Pr \left( Y_{t-1}^{1}=0|S_{t-2},V\right) >\Pr \left(
Y_{t-1}^{0}=0|S_{t-2},V\right) ,  \label{eq1b}
\end{equation}%
for all $V$. Assume that (\ref{eq1a}) holds. By Assumption \ref{Single} this
implies that
\begin{equation*}
\Pr (Y_{t-1}^{1}=0,Y_{t-1}^{0}=1|S_{i,t-2},V)=0,
\end{equation*}%
so that
\begin{align*}
\Pr (Y_{t-1}^{1}=0|S_{t-2},V)& =\Pr
(Y_{t-1}^{1}=0,Y_{t-1}^{0}=0|S_{t-2},V)+\Pr
(Y_{t-1}^{1}=0,Y_{t-1}^{0}=1|S_{t-2},V) \\
& =\Pr (Y_{t-1}^{1}=0,Y_{t-1}^{0}=0|S_{t-2},V).
\end{align*}%
Because this holds for all $V$ we omit $V$ in the sequel. Because
Assumptions \ref{MTR-gen} and \ref{Single} hold for all $t$, it follows from
this equation by recursion that
\begin{equation*}
\Pr (Y_{t-1}^{1}=0,\ldots ,Y_{1}^{1}=0,Y_{t-1}^{0}=0,\ldots
,Y_{1}^{0}=0)=\prod_{s=1}^{t-1}\Pr (Y_{s}^{1}=0|\overline{Y}_{s-1}^{1}=0),
\end{equation*}%
so that
\begin{equation*}
p_{t-1}(0,0)=\prod_{s=1}^{t-1}\Pr (Y_{s}^{1}=0|\overline{Y}%
_{s-1}^{1}=0)=\prod_{s=1}^{t-1}\Pr (Y_{s}=0|\overline{Y}_{s-1}=0,D=1).
\end{equation*}%
If Assumption \ref{MTR-gen} holds with (\ref{eq1b}), then
\begin{equation*}
p_{t-1}(0,0)=\prod_{s=1}^{t-1}\Pr (Y_{s}^{0}=0|\overline{Y}%
_{s-1}^{0}=0)=\prod_{s=1}^{t-1}\Pr (Y_{s}=0|\overline{Y}_{s-1}=0,D=0).
\end{equation*}%
We conclude that
\begin{equation*}
p_{t-1}(0,0)\geq \min \left \{ \prod_{s=1}^{t-1}\Pr (Y_{s}=0|\overline{Y}%
_{s-1}=0,D=1),\prod_{s=1}^{t-1}\Pr (Y_{s}=0|\overline{Y}_{s-1}=0,D=0)\right%
\} =
\end{equation*}%
\begin{equation*}
\min \left \{ \Pr (\overline{Y}_{t-1}=0|D=1),\Pr (\overline{Y}%
_{t-1}=0|D=0)\right \} .
\end{equation*}%
As noted below Theorem \ref{ATETSBoundsMTR-t} the bounds simplifies in an
obvious way if we have prior knowledge of the direction of the effect of the
treatment.

Next, upon substitution of this lower bound on $p_{t-1}(0,0)$ into (\ref%
{ATETSt2}) and because the probability $\mathbb{E}\left[ Y_{t}^{0}|\overline{%
Y}_{t-1}^{1}=0\right] $ is bounded by zero and one we have

\begin{align*}
& \max \Bigg \{0\;,\frac{[\Pr (Y_{t}=1|\overline{Y}_{t-1}=0,D=0)-1]\Pr
\left( \overline{Y}_{t-1}=0|D=0\right) }{\Pr (\overline{Y}_{t-1}=0|D=1)} \\
& \quad \qquad \qquad \qquad +\frac{\min \left \{ \Pr (\overline{Y}%
_{t-1}=0|D=1),\Pr (\overline{Y}_{t-1}=0|D=0)\right \} }{\Pr (\overline{Y}%
_{t-1}=0|D=1)}\Bigg \}.
\end{align*}

\begin{equation}
\leq \mathbb{E}\left[ Y_{t}^{0}|\overline{Y}_{t-1}^{1}=0\right] \leq
\label{teqsolmaxminMTR}
\end{equation}%
\begin{align*}
& \min \Bigg \{1,\;1+\frac{\Pr (Y_{t}=1|\overline{Y}_{t-1}=0,D=0)\Pr \left(
\overline{Y}_{t-1}=0|D=0\right) }{\Pr (\overline{Y}_{t-1}=0|D=1)} \\
& \qquad \qquad \qquad \quad -\frac{\min \left \{ \Pr (\overline{Y}%
_{t-1}=0|D=1),\Pr (\overline{Y}_{t-1}=0|D=0)\right \} }{\Pr (\overline{Y}%
_{t-1}=0|D=1)}\Bigg \},
\end{align*}%
Finally, we combine these bounds with the point-identified $\mathbb{E}%
[Y_{t}^{1}|\overline{Y}_{t-1}^{1}=0]$ to obtain bounds on $\mathrm{ATETS}%
_{t} $.\newline

\noindent \textbf{Proof of Theorem \ref{ATETS-Th-PCO-2}}\newline

\noindent As above, under Assumption 1 $\mathbb{E}[Y_{t}^{1}|\overline{Y}%
_{t-1}^{1}=0]=\Pr (Y_{t}=1|\overline{Y}_{t-1}=0,D=1),$ so that if $\Pr (%
\overline{Y}_{t-1}=0|D=1)>0$ then $\mathbb{E}[Y_{t}^{1}|\overline{Y}%
_{t-1}^{1}=0]$ is point-identified, and if $\Pr (\overline{Y}_{t-1}=0|D=1)=0$
then $\mathrm{ATETS}_{t}$ is not defined.

Next, we have for the counterfactual transition probability
\begin{equation}
\mathbb{E}\left[ Y_{t}^{0}|\overline{Y}_{t-1}^{1}=0\right] =\frac{%
p_{t}^{0}(1|0,0)p_{t-1}(0,0)+p_{t}^{0}(1|0,\neq 0)p_{t-1}(0,\neq 0)}{%
p_{t-1}(0,0)+p_{t-1}(0,\neq 0)}.  \label{teqbas00_PCO}
\end{equation}%
The expression on the right-hand side is increasing in $p_{t}^{0}(1|0,\neq
0) $. By Assumption \ref{PCO} we have the restriction $p_{t}^{0}(1|0,\neq
0)\geq p_{t}^{0}(1|0,0)$. Then the upper bound is obtained by setting $%
p_{t}^{0}(1|0,\neq 0)=1$ and lower bound by setting $p_{t}^{0}(1|0,\neq
0)=p_{t}^{0}(1|0,0)$:
\begin{equation*}
p_{t}^{0}(1|0,0)\leq \mathbb{E}\left[ Y_{t}^{0}|\overline{Y}_{t-1}^{1}=0%
\right] \leq \frac{p_{t}^{0}(1|0,0)p_{t-1}(0,0)+p_{t-1}(0,\neq 0)}{%
p_{t-1}(0,0)+p_{t-1}(0,\neq 0)}.
\end{equation*}%
By Assumption 1 and the law of total probability we have using similar
reasoning as for Theorem 1:%
\begin{equation}
\Pr (Y_{t}=1,\overline{Y}%
_{t-1}=0|D=0)=p_{t}^{0}(1|0,0)p_{t-1}(0,0)+p_{t}^{0}(1|\neq 0,0)p_{t-1}(\neq
0,0)  \label{PCOn}
\end{equation}%
Solving for $p_{t}^{0}(1|0,0)$ gives
\begin{equation*}
p_{t}^{0}(1|0,0)=\frac{\Pr (Y_{t}=1,\overline{Y}_{t-1}=0|D=0)-p_{t}^{0}(1|%
\neq 0,0)p_{t-1}(\neq 0,0)}{p_{t-1}(0,0)}
\end{equation*}%
and upon substitution%
\begin{equation*}
\frac{\Pr (Y_{t}=1,\overline{Y}_{t-1}=0|D=0)-p_{t}^{0}(1|\neq
0,0)p_{t-1}(\neq 0,0)}{p_{t-1}(0,0)}\leq \mathbb{E}\left[ Y_{t}^{0}|%
\overline{Y}_{t-1}^{1}=0\right] \leq
\end{equation*}%
\begin{equation*}
\frac{\Pr (Y_{t}=1,\overline{Y}_{t-1}=0|D=0)-p_{t}^{0}(1|\neq
0,0)p_{t-1}(\neq 0,0)+p_{t-1}(0,\neq 0)}{p_{t-1}(0,0)+p_{t-1}(0,\neq 0)}
\end{equation*}%
Both the lower and upper bound is decreasing in $p_{t}^{0}(1|\neq 0,0)$. By
Assumption \ref{PCO} we have the restriction $p_{t}^{0}(1|\neq 0,0)\geq
p_{t}^{0}(1|0,0)$. Therefore the lower bound is obtained by setting $%
p_{t}^{0}(1|\neq 0,0)$ at 1. The upper bound is obtained by setting $%
p_{t}^{0}(1|\neq 0,0)=p_{t}^{0}(1|0,0)$, upon substitution into (\ref{PCOn})
this implies that
\begin{equation*}
p_{t}^{0}(1|\neq 0,0)=p_{t}^{0}(1|0,0)=\Pr (Y_{t}=1|\overline{Y}%
_{t-1}=0,D=0).
\end{equation*}%
Then,%
\begin{equation*}
\frac{\Pr (Y_{t}=1,\overline{Y}_{t-1}=0|D=0)-p_{t-1}(\neq 0,0)}{p_{t-1}(0,0)}%
\leq \mathbb{E}\left[ Y_{t}^{0}|\overline{Y}_{t-1}^{1}=0\right] \leq
\end{equation*}%
\begin{equation*}
\frac{\Pr (Y_{t}=1,\overline{Y}_{t-1}=0|D=0)-\Pr (Y_{t}=1|\overline{Y}%
_{t-1}=0,D=0)p_{t-1}(\neq 0,0)+p_{t-1}(0,\neq 0)}{p_{t-1}(0,0)+p_{t-1}(0,%
\neq 0)}
\end{equation*}%
Because
\begin{equation*}
\Pr (\overline{Y}_{t-1}=0|D=1)=p_{t-1}(0,0)+p_{t-1}(0,\neq 0)
\end{equation*}%
\begin{equation*}
\Pr (\overline{Y}_{t-1}=0|D=0)=p_{t-1}(0,0)+p_{t-1}(\neq 0,0)
\end{equation*}%
we have
\begin{equation}
\frac{\Pr (Y_{t}=1,\overline{Y}_{t-1}=0|D=0)-\Pr (\overline{Y}%
_{t-1}=0|D=0)+p_{t-1}(0,0)}{p_{t-1}(0,0)}  \label{ATETSt2_PCO}
\end{equation}%
\begin{equation*}
\leq \mathbb{E}\left[ Y_{t}^{0}|\overline{Y}_{t-1}^{1}=0\right] \leq \frac{%
\lbrack \Pr (Y_{t}=1|\overline{Y}_{t-1}=0,D=0)-1]p_{t-1}(0,0)}{\Pr (%
\overline{Y}_{t-1}=0|D=1)}+1.
\end{equation*}%
The lower bound is increasing and the upper bound decreasing in $%
p_{t-1}(0,0) $. Assumption \ref{PCO} also improves on the Bonferroni
inequality for $p_{t-1}(0,0)$. We have
\begin{equation*}
p_{t-1}(0,0)=\prod_{s=1}^{t-1}\Pr (Y_{s}^{1}=0,Y_{s}^{0}=0|S_{s-1}).
\end{equation*}%
By the Bonferroni inequality and the results above
\begin{equation*}
\Pr (Y_{s}^{1}=0,Y_{s}^{0}=0|S_{s-1})\geq \max \{1-\Pr
(Y_{s}^{1}=1|S_{s-1})-\Pr (Y_{s}^{0}=1|S_{s-1}),0\} \geq
\end{equation*}%
\begin{equation*}
\max \{1-\Pr (Y_{s}=1|\overline{Y}_{s-1}=0,D=1)-\Pr (Y_{s}=1|\overline{Y}%
_{s-1}=0,D=0),0\}=
\end{equation*}%
\begin{equation*}
\max \{ \Pr (Y_{s}=0|\overline{Y}_{s-1}=0,D=1)+\Pr (Y_{s}=0|\overline{Y}%
_{s-1}=0,D=0)-1,0\},
\end{equation*}%
so that
\begin{equation}
p_{t-1}(0,0)\geq \prod_{s=1}^{t-1}\max \{ \Pr (Y_{s}=0|\overline{Y}%
_{s-1}=0,D=1)+\Pr (Y_{s}=0|\overline{Y}_{s-1}=0,D=0)-1,0\}.  \label{eqATETSs}
\end{equation}%
We compare this to the lower bound
\begin{equation*}
\max \left \{ \prod_{s=1}^{t-1}\Pr (Y_{s}=0|\overline{Y}_{s-1}=0,D=1)+%
\prod_{s=1}^{t-1}\Pr (Y_{s}=0|\overline{Y}_{s-1}=0,D=0)-1,0\right \}
\end{equation*}%
that we obtained in the proof of Theorem 1. First, if there is an $1\leq
s^{\prime }\leq t-1$ so that
\begin{equation*}
\Pr (Y_{s^{\prime }}=0|\overline{Y}_{s^{\prime }-1}=0,D=1)+\Pr (Y_{s^{\prime
}}=0|\overline{Y}_{s^{\prime }-1}=0,D=0)-1<0,
\end{equation*}%
then
\begin{equation*}
\prod_{s=1}^{t-1}\Pr (Y_{s}=0|\overline{Y}_{s-1}=0,D=1)+\prod_{s=1}^{t-1}\Pr
(Y_{s}=0|\overline{Y}_{s-1}=0,D=0)-1=
\end{equation*}%
\begin{equation*}
\Pr (Y_{s^{\prime }}=0|\overline{Y}_{s^{\prime }-1}=0,D=1)\prod_{s=1,s\neq
s^{\prime }}^{t-1}\Pr (Y_{s}=0|\overline{Y}_{s-1}=0,D=1)+
\end{equation*}%
\begin{equation*}
\Pr (Y_{s^{\prime }}=0|\overline{Y}_{s^{\prime }-1}=0,D=1)\prod_{s=1,s\neq
s^{\prime }}^{t-1}\Pr (Y_{s}=0|\overline{Y}_{s-1}=0,D=0)-1<0
\end{equation*}%
so that if the new lower bound is 0, so is the previous one. Finally, if for
all $s=1,\ldots ,t-1$
\begin{equation*}
\Pr (Y_{s}=0|\overline{Y}_{s-1}=0,D=1)+\Pr (Y_{s}=0|\overline{Y}%
_{s-1}=0,D=0)-1>0,
\end{equation*}%
then
\begin{equation*}
\prod_{s=1}^{t-1}\left[ \Pr (Y_{s}=0|\overline{Y}_{s-1}=0,D=1)+\Pr (Y_{s}=0|%
\overline{Y}_{s-1}=0,D=0)-1\right] \geq
\end{equation*}%
\begin{equation*}
\prod_{s=1}^{t-1}\Pr (Y_{s}=0|\overline{Y}_{s-1}=0,D=1)+\prod_{s=1}^{t-1}\Pr
(Y_{s}=0|\overline{Y}_{s-1}=0,D=0)-1.
\end{equation*}

If $\Pr (Y_{s}=0|\overline{Y}_{s-1}=0,D=1)+\Pr (Y_{s}=0|\overline{Y}%
_{s-1}=0,D=0)-1\leq 0$ for some $s\leq t$ the lower bound on $p_{t-1}(0,0)$
is 0. In that case the lower bound in (\ref{ATETSt2_PCO}) is non-positive
and the upper bound is greater than or equal to 1 so that $0\leq \mathbb{E}%
\left[ Y_{t}^{0}|\overline{Y}_{t-1}^{1}=0\right] \leq 1$.

If $\Pr (Y_{s}=0|\overline{Y}_{s-1}=0,\overline{D}_{s}=1)+\Pr (Y_{s}=0|%
\overline{Y}_{s-1}=0,D=0)-1>0$ for all $s=1,\ldots ,t-1$ we have upon
substitution of the lower bound on $p_{t-1}(0,0)$ in (\ref{eqATETSs}) into (%
\ref{ATETSt2_PCO}) and because the probability $\mathbb{E}\left[ Y_{t}^{0}|%
\overline{Y}_{t-1}^{1}=0\right] $ is bounded by zero,%
\begin{equation*}
\max \left \{ 0,\frac{(\Pr (Y_{t}=1|\overline{Y}_{t-1}=0,D=0)-1)\Pr \left(
\overline{Y}_{t-1}=0|D=0\right) }{\prod_{s=1}^{t-1}[\Pr (Y_{s}=0|\overline{Y}%
_{s-1}=0,D=1)+\Pr (Y_{s}=0|\overline{Y}_{s-1}=0,D=0)-1]}+1\right \}
\end{equation*}%
\begin{equation}
\leq \mathbb{E}\left[ Y_{t}^{0}|\overline{Y}_{t-1}^{1}=0\right] \leq 1-\frac{%
1-\Pr (Y_{t}=1|\overline{Y}_{t-1}=0,D=0)}{\Pr (\overline{Y}_{t-1}=0|D=1)}%
\cdot  \label{teqsolmaxmin11_PCO}
\end{equation}%
\begin{equation*}
\qquad \qquad \cdot \prod_{s=1}^{t-1}[\Pr (Y_{s}=0|\overline{Y}%
_{s-1}=0,D=1)+\Pr (Y_{s}=0|\overline{Y}_{s-1}=0,D=0)-1]\text{.}
\end{equation*}%
Finally, we combine these bounds with the point-identified $\mathbb{E}%
[Y_{t}^{1}|\overline{Y}_{t-1}^{1}=0]$ to obtain bounds on $\mathrm{ATETS}%
_{t} $.\newline

\noindent \textbf{Proof of Theorem \ref{ATETSBoundsMTRPCO-t}}\newline

\noindent Using similar reasoning as for the proof of Theorem \ref%
{ATETS-Th-PCO-2} we have under Assumptions 1 and \ref{PCO}:%
\begin{equation*}
\mathbb{E}[Y_{t}^{1}|\overline{Y}_{t-1}^{1}=0]=\Pr (Y_{t}=1|\overline{Y}%
_{t-1}=0,D=1)
\end{equation*}%
and%
\begin{equation*}
\frac{\Pr (Y_{t}=1,\overline{Y}_{t-1}=0|D=0)-\Pr (\overline{Y}%
_{t-1}=0|D=0)+p_{t-1}(0,0)}{p_{t-1}(0,0)}
\end{equation*}%
\begin{equation*}
\leq \mathbb{E}\left[ Y_{t}^{0}|\overline{Y}_{t-1}^{1}=0\right] \leq \frac{%
\lbrack \Pr (Y_{t}=1|\overline{Y}_{t-1}=0,D=0)-1]p_{t-1}(0,0)}{\Pr (%
\overline{Y}_{t-1}=0|D=1)}+1.
\end{equation*}%
The lower bound on $\mathbb{E}\left[ Y_{t}^{0}|\overline{Y}_{t-1}^{1}=0%
\right] $ is increasing and the upper bound on $\mathbb{E}\left[ Y_{t}^{0}|%
\overline{Y}_{t-1}^{1}=0\right] $ is decreasing in $p_{t-1}(0,0)$. By the
proof of Theorem \ref{ATETSBoundsMTR-t} we have under Assumptions \ref%
{MTR-gen} and \ref{Single}
\begin{equation*}
p_{t-1}(0,0)\geq \min \left \{ \Pr (\overline{Y}_{t-1}=0|D=1),\Pr (\overline{%
Y}_{t-1}=0|D=0)\right \} ,
\end{equation*}%
so that

\begin{equation*}
\max \left \{ 0,\frac{(\Pr (Y_{t}=1|\overline{Y}_{t-1}=0,D=0)-1)\Pr \left(
\overline{Y}_{t-1}=0|D=0\right) }{\min \left \{ \Pr (\overline{Y}%
_{t-1}=0|D=1),\Pr (\overline{Y}_{t-1}=0|D=0)\right \} }+1\right \} \leq
\mathbb{E}\left[ Y_{t}^{0}|\overline{Y}_{t-1}^{1}=0\right] \leq
\end{equation*}%
\begin{equation*}
\frac{1-\Pr (Y_{t}=1|\overline{Y}_{t-1}=0,D=0)}{\Pr (\overline{Y}%
_{t-1}=0|D=1)}\times \min \left \{ \Pr (\overline{Y}_{t-1}=0|D=1),\Pr (%
\overline{Y}_{t-1}=0|D=0)\right \} +1.
\end{equation*}%
Together with the results for $\mathbb{E}[Y_{t}^{1}|\overline{Y}%
_{t-1}^{1}=0] $ this gives the bounds.\pagebreak

\addcontentsline{toc}{section}{Appendix B}

\section*{Appendix B: Heterogenous effects (for online publication only)}

\renewcommand{\theequation}{B.\arabic{equation}} \setcounter{equation}{0}

\begin{table}[h!]
\caption{Bounds on $ATETS^{1,0}$ for the Illinois job bonus experiment.
Heterogenous effects for blacks and non-blacks}
\label{t:bounds_race}\centering{\small
\begin{tabularx}{\textwidth}{lzzzzzzzz}
\toprule
&     \multicolumn{8}{c}{\textbf{Panel A: Blacks}}    \\
&     \multicolumn{4}{c}{No assumption bounds} & \multicolumn{4}{c}{MTR+CS+PCO}   \\
\cmidrule(r){2-5}
\cmidrule(l){6-9}
& Lower-CI &  LB & UB & Upper-CI & Lower-CI &  LB & UB & Upper-CI \\
& (1) & (2) & (3) & (4) & (1) & (2) & (3) & (4) \\
\hline
   Week &        \\
       1-2   &-0.006& ~0.021 &0.021 &0.049    &~0.000& ~0.021 &0.021 &0.043              \\
       3-4   &-0.124& -0.111& 0.059& 0.071    &-0.012& ~0.005 &0.028 &0.044       \\
       5-6   &-0.180& -0.167& 0.058& 0.070    &-0.004& ~0.015 &0.043 &0.062        \\
       7-8   &-0.243& -0.230& 0.044& 0.057    &-0.007& ~0.010 &0.044 &0.061       \\
       9-10  &-0.290& -0.277& 0.048& 0.060    &-0.005& ~0.012 &0.048 &0.064        \\
       11-12 &-0.352& -0.342& 0.030& 0.040    &-0.013& ~0.001 &0.030 &0.044       \\
       13-14 &-0.395& -0.384& 0.032& 0.043    &-0.012& ~0.002 &0.032 &0.045      \\
       15-16 &-0.449& -0.439& 0.025& 0.035    &-0.020& -0.007& 0.025& 0.037      \\
       17-18 &-0.496& -0.485& 0.028& 0.039    &-0.021& -0.007& 0.028& 0.042      \\
       19-20 &-0.532& -0.520& 0.037& 0.049    &-0.007& ~0.010 &0.037 &0.053      \\
       21-22 &-0.605& -0.596& 0.019& 0.029    &-0.028& -0.016& 0.019& 0.031     \\
       23-24 &-0.635& -0.623& 0.039& 0.051    &-0.011& ~0.006 &0.039 &0.055      \\

\hline
&     \multicolumn{4}{c}{PCO [C]} & \multicolumn{4}{c}{MTR+CS+PCO [D]}   \\
\cmidrule(r){2-5}
\cmidrule(l){6-9}
&     \multicolumn{8}{c}{\textbf{Panel B: Non-blacks}}    \\
&     \multicolumn{4}{c}{No assumption bounds} & \multicolumn{4}{c}{MTR+CS+PCO}   \\
& (1) & (2) & (3) & (4) & (1) & (2) & (3) & (4) \\
\hline
   Week &        \\
       1-2    & ~0.005  &~0.022   & 0.022  &0.040      &~0.009 &~0.022& 0.022& 0.035            \\
       3-4    & -0.158 & -0.148 & 0.106  &0.116        &~0.002 &~0.016& 0.040& 0.053     \\
       5-6    & -0.293 & -0.284 & 0.080  &0.090        &-0.010 &~0.003& 0.044& 0.058     \\
       7-8    & -0.392 & -0.382 & 0.090  &0.100        &~0.003 &~0.017& 0.062& 0.076     \\
       9-10   & -0.523 & -0.513 & 0.077  &0.087        &~0.001 &~0.011& 0.074& 0.086      \\
       11-12  & -0.639 & -0.629 & 0.075  &0.085        &~0.006 &~0.015& 0.075& 0.087     \\
       13-14  & -0.773 & -0.763 & 0.066  &0.076        &-0.006 &~0.004& 0.066& 0.077   \\
       15-16  & -0.889 & -0.879 & 0.062  &0.071        &~0.003 &~0.013& 0.062& 0.072     \\
       17-18  & -0.991 & -0.942 & 0.058  &0.068        &~ 0.000&~0.010& 0.058& 0.069    \\
       19-20  & -1.002 & -0.946 & 0.054  &0.064        &-0.013 &-0.003& 0.054& 0.064  \\
       21-22  & -1.002 & -0.940 & 0.060  &0.071        &-0.024 &-0.013& 0.060& 0.073  \\
       23-24  & -1.008 & -0.936 & 0.064  &0.076        &-0.011 &0.001 &0.064 &0.079   \\
\bottomrule
\multicolumn{9}{p{0.98\textwidth}}{\footnotesize Notes: CI is 95\% confidence intervals. Variances and covariances used to obtain the CI are estimated using bootstrap (399 replications).}
\end{tabularx}
}
\end{table}

\clearpage \newpage

\begin{table}[h]
\caption{Bounds on $ATETS^{1,0}$ for the Illinois job bonus experiment.
Heterogenous effects for low and high income workers}
\label{t:bounds_income}\centering{\small
\begin{tabularx}{\textwidth}{lzzzzzzzz}
\toprule
&     \multicolumn{8}{c}{\textbf{Panel A: Below median income}}    \\
&     \multicolumn{4}{c}{No assumption bounds} & \multicolumn{4}{c}{MTR+CS+PCO}   \\
\cmidrule(r){2-5}
\cmidrule(l){6-9}
& Lower-CI &  LB & UB & Upper-CI & Lower-CI &  LB & UB & Upper-CI \\
& (1) & (2) & (3) & (4) & (1) & (2) & (3) & (4) \\
\hline
   Week &        \\
       1-2   &-0.006& ~0.016& 0.016& 0.038     &-0.001 &~0.016 &0.016 &0.032              \\
       3-4   &-0.172& -0.161& 0.089& 0.100     &-0.002 &~0.013 &0.030 &0.044       \\
       5-6   &-0.285& -0.274& 0.067& 0.078     &-0.012 &~0.002 &0.033 &0.047        \\
       7-8   &-0.355& -0.344& 0.080& 0.091     &0.009  &~0.024 &0.057 &0.072        \\
       9-10  &-0.461& -0.450& 0.067& 0.078     &0.003  &~0.014 &0.067 &0.080         \\
       11-12 &-0.568& -0.558& 0.050& 0.060     &-0.010 &~0.000 &0.050 &0.062      \\
       13-14 &-0.644& -0.634& 0.047& 0.056     &-0.001 &~0.008 &0.047 &0.058      \\
       15-16 &-0.727& -0.717& 0.041& 0.051     &-0.009 &~0.001 &0.041 &0.052      \\
       17-18 &-0.802& -0.792& 0.040& 0.050     &-0.008 &~0.002 &0.040 &0.051      \\
       19-20 &-0.883& -0.869& 0.049& 0.063     &-0.008 &~0.003 &0.049 &0.062      \\
       21-22 &-1.021& -0.959& 0.041& 0.053     &-0.024 &-0.014 &0.041 &0.053    \\
       23-24 &-1.015& -0.958& 0.042& 0.053     &-0.021 &-0.010 &0.042 &0.055     \\

\hline
&     \multicolumn{4}{c}{PCO [C]} & \multicolumn{4}{c}{MTR+CS+PCO [D]}   \\
\cmidrule(r){2-5}
\cmidrule(l){6-9}
&     \multicolumn{8}{c}{\textbf{Panel B: Above median income}}    \\
&     \multicolumn{4}{c}{No assumption bounds} & \multicolumn{4}{c}{MTR+CS+PCO}   \\
& (1) & (2) & (3) & (4) & (1) & (2) & (3) & (4) \\
\hline
   Week &        \\
       1-2   & ~0.010 &~0.028& 0.028& 0.047            &~0.014 &~0.028& 0.028& 0.043             \\
       3-4   & -0.125 &-0.113& 0.099& 0.111            &~0.001 &~0.015& 0.045& 0.059      \\
       5-6   & -0.238 &-0.227& 0.081& 0.092            &-0.002 &~0.012& 0.059& 0.073      \\
       7-8   & -0.342 &-0.331& 0.076& 0.087            &-0.003 &~0.008& 0.068& 0.000     \\
       9-10  & -0.447 &-0.435& 0.070& 0.082            &-0.002 &~0.010& 0.070& 0.084      \\
       11-12 & -0.535 &-0.524& 0.074& 0.086            &~0.012 &~0.024& 0.074& 0.089      \\
       13-14 & -0.666 &-0.653& 0.066& 0.078            &-0.012 &~0.000& 0.066& 0.080   \\
       15-16 & -0.772 &-0.760& 0.060& 0.072            &~0.003 &~0.015& 0.060& 0.072      \\
       17-18 & -0.881 &-0.870& 0.058& 0.070            &-0.003 &~0.009& 0.058& 0.070     \\
       19-20 & -1.008 &-0.952& 0.048& 0.059            &-0.010 &~0.000& 0.048& 0.058    \\
       21-22 & -1.010 &-0.947& 0.053& 0.065            &-0.024 &-0.013& 0.053& 0.065   \\
       23-24 & -0.999 &-0.929& 0.071& 0.086            &~0.004 &~0.018& 0.071& 0.086     \\
\bottomrule
\multicolumn{9}{p{0.98\textwidth}}{\footnotesize Notes: CI is 95\% confidence intervals. Variances and covariances used to obtain the CI are estimated using bootstrap (399 replications).}
\end{tabularx}
}
\end{table}

\clearpage \newpage \addcontentsline{toc}{section}{Appendix C}

\section*{Appendix C: Average treatment effect on survivors (for online
publication only)}

\renewcommand{\theequation}{C.\arabic{equation}} \setcounter{equation}{0}

In this appendix we consider the average effect when averaging over the
subpopulation of individuals who would have survived until $t$ under both
treatment and no-treatment. We call this average effect the Average
Treatment Effect on Survivors, $\mathrm{ATES}_{t}$:

\begin{definition}
Average Treatment Effect on Survivors (ATES)%
\begin{equation}
\mathrm{ATES}_{t}=\mathbb{E}\left( Y_{t}^{1}|\overline{Y}%
_{t-1}^{1}=0,Y_{t-1}^{0}=0\right) -\mathbb{E}\left( Y_{t}^{0}|\overline{Y}%
_{t-1}^{1}=0,Y_{t-1}^{0}=0\right)  \notag
\end{equation}
\end{definition}

The bounds for $\mathrm{ATES}_{t}$ are given in Theorem \ref{BasBounds-t}.

\begin{theorem}[Bounds on ATES]
\label{BasBounds-t} Suppose that Assumption \ref{AssRand3} holds. If $\Pr
\left( \overline{Y}_{t-1}=0|D=1\right) +\Pr \left( \overline{Y}%
_{t-1}=0|D=0\right) -1\leq 0$, then $\mathrm{ATES}_{t}$ is not defined.

If $\Pr \left( \overline{Y}_{t-1}=0|D=1\right) +\Pr \left( \overline{Y}%
_{t-1}=0|D=0\right) -1>0$, then we have the following sharp bounds%
\begin{equation}
\max \left \{ 0,\frac{\Pr (Y_{t}=1,\overline{Y}_{t-1}=0|D=1)+\Pr \left(
\overline{Y}_{t-1}=0|D=0\right) -1}{\Pr \left( \overline{Y}%
_{t-1}=0|D=1\right) +\Pr \left( \overline{Y}_{t-1}=0|D=0\right) -1}\right \}
-  \notag
\end{equation}%
\begin{equation*}
\min \left \{ 1,\frac{\Pr (Y_{t}=1,\overline{Y}_{t-1}=0|D=0)}{\Pr \left(
\overline{Y}_{t-1}=0|D=0\right) +\Pr \left( \overline{Y}_{t-1}=0|D=1\right)
-1}\right \} \leq \mathrm{ATES}_{t}\leq
\end{equation*}%
\begin{equation*}
\min \left \{ 1,\frac{\Pr (Y_{t}=1,\overline{Y}_{t-1}=0|D=1)}{\Pr \left(
\overline{Y}_{t-1}=0|D=1\right) +\Pr \left( \overline{Y}_{t-1}=0|D=0\right)
-1}\right \} -
\end{equation*}%
\begin{equation*}
\max \left \{ 0,\frac{\Pr (Y_{t}=1,\overline{Y}_{t-1}=0|D=0)+\Pr \left(
\overline{Y}_{t-1}=0|D=1\right) -1}{\Pr \overline{Y}_{t-1}=0|D=1)+\Pr \left(
\overline{Y}_{t-1}=0|D=1\right) -1}\right \} .
\end{equation*}%
\noindent
\end{theorem}

\noindent \textbf{Proof:} First, consider bounds on $\mathbb{E}\left[
Y_{t}^{1}|\overline{Y}_{t-1}^{1}=0,\overline{Y}_{t-1}^{0}=0\right]
=p_{t}^{1}(1|0,0)$. By Assumption 2%
\begin{equation*}
\Pr (Y_{t}=1,\overline{Y}_{t-1}=0|D=1)=\Pr (Y_{t}^{1}=1,\overline{Y}%
_{t-1}^{1}=0).
\end{equation*}%
By the law of total probability
\begin{equation*}
\Pr (Y_{t}^{1}=1,\overline{Y}%
_{t-1}^{1}=0)=p_{t}^{0}(1|0,0)p_{t-1}(0,0)+p_{t}^{0}(1|0,\neq
0)p_{t-1}(0,\neq 0)
\end{equation*}%
Therefore,
\begin{equation*}
\Pr (Y_{t}=1,\overline{Y}%
_{t-1}=0|D=1)=p_{t}^{0}(1|0,0)p_{t-1}(0,0)+p_{t}^{0}(1|0,\neq
0)p_{t-1}(0,\neq 0)
\end{equation*}%
Solving for $p_{t}^{1}(1|0,0)=\mathbb{E}\left[ Y_{t}^{1}|\overline{Y}%
_{t-1}^{1}=0,\overline{Y}_{t-1}^{0}=0\right] $ gives
\begin{equation*}
\mathbb{E}\left[ Y_{t}^{1}|\overline{Y}_{t-1}^{1}=0,\overline{Y}_{t-1}^{0}=0%
\right] =\frac{\Pr (Y_{t}=1,\overline{Y}_{t-1}=0|D=1)-p_{t}^{0}(1|0,\neq
0)p_{t-1}(0,\neq 0)}{p_{t-1}(0,0)}
\end{equation*}%
The expression on the right-hand side is decreasing in $p_{t}^{0}(1|0,\neq
0) $. The lower bound is obtained by setting $p_{t}^{0}(1|0,\neq 0)$ at 1
and the upper bound by setting $p_{t}^{0}(1|0,\neq 0)$ at 0.
\begin{equation*}
\frac{\Pr (Y_{t}=1,\overline{Y}_{t-1}=0|D=1)-p_{t-1}(0,\neq 0)}{p_{t-1}(0,0)}
\end{equation*}%
\begin{equation*}
\leq \mathbb{E}\left[ Y_{t}^{1}|\overline{Y}_{t-1}^{1}=0,\overline{Y}%
_{t-1}^{0}=0\right] \leq \frac{\Pr (Y_{t}=1,\overline{Y}_{t-1}=0|D=1)}{%
p_{t-1}(0,0)}.
\end{equation*}%
Because
\begin{equation*}
\Pr (\overline{Y}_{t-1}=0|D=1)=p_{t-1}(0,0)+p_{t-1}(0,\neq 0)
\end{equation*}%
we have
\begin{equation*}
\frac{\Pr (Y_{t}=1,\overline{Y}_{t-1}=0|D=1)-\Pr (\overline{Y}%
_{t-1}=0|D=1)+p_{t-1}(0,0)}{p_{t-1}(0,0)}
\end{equation*}%
\begin{equation*}
\leq \mathbb{E}\left[ Y_{t}^{1}|\overline{Y}_{t-1}^{1}=0,\overline{Y}%
_{t-1}^{0}=0\right] \leq \frac{\Pr (Y_{t}=1,\overline{Y}_{t-1}=0|D=1)}{%
p_{t-1}(0,0)}.
\end{equation*}%
The upper bound is decreasing and the lower bound is increasing in $%
p_{t-1}(0,0)$. From the proof of theorem \ref{ATETSBasBounds-t} we have
\begin{equation*}
p_{t-1}(0,0)\geq \max \left \{ \Pr \left( \overline{Y}_{t-1}=0|D=1\right)
+\Pr \left( \overline{Y}_{t-1}=0|D=0\right) -1,0\right \} .
\end{equation*}%
If $\Pr \left( \overline{Y}_{t-1}=0|D=1\right) +\Pr \left( \overline{Y}%
_{t-1}=0|D=0\right) -1>0$ then we are sure that there are survivors in both
treatment arms. Upon substitution of this lower bound%
\begin{equation*}
\frac{\Pr (Y_{t}=1,\overline{Y}_{t-1}=0|D=1)+\Pr \left( \overline{Y}%
_{t-1}=0|D=0\right) -1}{\Pr \left( \overline{Y}_{t-1}=0|D=1\right) +\Pr
\left( \overline{Y}_{t-1}=0|D=0\right) -1}
\end{equation*}%
\begin{equation*}
\leq \mathbb{E}\left[ Y_{t}^{1}|\overline{Y}_{t-1}^{1}=0,\overline{Y}%
_{t-1}^{0}=0\right] \leq \frac{\Pr (Y_{t}=1,\overline{Y}_{t-1}=0|D=1)}{\Pr
\left( \overline{Y}_{t-1}=0|D=1\right) +\Pr \left( \overline{Y}%
_{t-1}=0|D=0\right) -1}.
\end{equation*}%
By an analogous argument we have%
\begin{equation*}
\frac{\Pr (Y_{t}=1,\overline{Y}_{t-1}=0|D=0)+\Pr \left( \overline{Y}%
_{t-1}=0|D=1\right) -1}{\Pr \overline{Y}_{t-1}=0|D=1)+\Pr \left( \overline{Y}%
_{t-1}=0|D=0\right) -1}
\end{equation*}%
\begin{equation*}
\leq \mathbb{E}\left[ Y_{t}^{0}|\overline{Y}_{t-1}^{1}=0,\overline{Y}%
_{t-1}^{0}=0\right] \leq \frac{\Pr (Y_{t}=1,\overline{Y}_{t-1}=0|D=0)}{\Pr
\left( \overline{Y}_{t-1}=0|D=1\right) +\Pr \left( \overline{Y}%
_{t-1}=0|D=0\right) -1}.
\end{equation*}%
Substitution of these results for $\mathbb{E}\left[ Y_{t}^{1}|\overline{Y}%
_{t-1}^{1}=0,\overline{Y}_{t-1}^{0}=0\right] $ and $\mathbb{E}\left[
Y_{t}^{0}|\overline{Y}_{t-1}^{1}=0,\overline{Y}_{t-1}^{0}=0\right] $ and
because both probabilites are bounded by zero and one gives the bounds on $%
\mathrm{ATES}_{t}$.

\end{document}